\documentclass[12pt,reqno]{amsart}
\usepackage{fullpage}
\usepackage{times}
\usepackage{graphicx}

\usepackage{amsmath,amsthm,amsfonts,amscd,amssymb}
\usepackage{pst-node}
\usepackage{pst-plot}

\usepackage{a4}
\usepackage{amssymb}
\usepackage{array}
\usepackage{booktabs}
\usepackage{multirow}
\usepackage{hhline}

\newtheorem{thm}{Theorem}

\newtheorem{cor}[thm]{Corollary}
\newtheorem{lem}[thm]{Lemma}
\newtheorem{prop}[thm]{Proposition}
\newtheorem{question}[thm]{Question}

\theoremstyle{definition}
\newtheorem{ex}[thm]{Example}
\theoremstyle{definition}
\newtheorem{defn}[thm]{Definition}
\theoremstyle{definition}
\newtheorem{rem}[thm]{Remark}

\numberwithin{thm}{section}
\numberwithin{equation}{section}
\newcommand{\Real}{\mathbb R}

\newcommand{\norm}[1]{\left\Vert#1\right\Vert}

\newcommand{\set}[1]{\left\{#1\right\}}

\newcommand{\la}{\langle}
\newcommand{\ra}{\rangle}

\newcommand{\B}{\mathcal{B}}

\newcommand{\Comp}{\mathbb{C}}

\newcommand{\D}{\mathcal{D}}

\newcommand{\F}{\mathcal{F}}

\newcommand{\Hi}{\mathcal{H}}

\newcommand{\tor}{\mathbb{T}}

\newcommand{\W}{\mathcal{W}}

\newcommand{\z}{\mathbb{Z}}

\newcommand{\Om}{\Omega}
\newcommand{\om}{\omega}

\newcommand{\weyl}{{W}}

\newcommand{\mc}[1]{\mathcal #1}
\newcommand{\tr}{{\rm Tr}\,}

\newcommand{\id}{{\rm id}}

\newcommand{\be}{\begin{equation}}
\newcommand{\ee}{\end{equation}}
\newcommand{\bs}{\begin{split}}
\newcommand{\es}{\end{split}}

\newcommand{\ten}{\otimes}
\newcommand{\wh}[1]{\widehat{#1}}

\newcommand{\lm}{\lambda}
\newcommand{\vphi}{\varphi}

\newcommand{\bC}{\mathbb{C}}
\newcommand{\bN}{\mathbb{N}}
\newcommand{\bZ}{\mathbb{Z}}

\newcommand{\bR}{\mathbb{R}}
\newcommand{\bT}{\mathbb{T}}

\begin{document}
\title{Gaussian  quantum information over general quantum kinematical systems I: Gaussian states}

\author{Cedric Beny}
\address{Cedric Beny: Cortex Discovery GmbH}
\email{cedric.beny@gmail.com}

\author{Jason Crann}
\address{Jason Crann: School of Mathematics \& Statistics, Carleton University, Ottawa, ON, Canada H1S 5B6}
\email{jasoncrann@cunet.carleton.ca}

\author{Hun Hee Lee}
\address{Hun Hee Lee: Department of Mathematical Sciences and the Research Institute of Mathematics, Seoul National University, Gwanak-ro 1, Gwanak-gu, Seoul 08826, Republic of Korea}
\email{hunheelee@snu.ac.kr}

\author{Sang-Jun Park}
\address{Sang-Jun Park: Department of Mathematical Sciences, Seoul National University, Gwanak-ro 1, Gwanak-gu, Seoul 08826, Republic of Korea}
\email{psj05071@snu.ac.kr}

\author{Sang-Gyun Youn}
\address{Sang-Gyun Youn: Department of Mathematics Education, Seoul National University, Gwanak-ro 1, Gwanak-gu, Seoul 08826, Republic of Korea}
\email{s.youn@snu.ac.kr}


\begin{abstract} We develop a theory of Gaussian states over general quantum kinematical systems with finitely many degrees of freedom. The underlying phase space is described by a locally compact abelian (LCA) group $G$ with a symplectic structure determined by a 2-cocycle on $G$. We use the concept of Gaussian distributions on LCA groups in the sense of Bernstein to define Gaussian states and completely characterize Gaussian states over 2-regular LCA groups of the form $G= F\times\widehat{F}$ endowed with a canonical normalized 2-cocycle. This covers, in particular, the case of $n$-bosonic modes, $n$-qudit systems with odd $d\ge 3$, and $p$-adic quantum systems. Our characterization reveals a topological obstruction to Gaussian state entanglement when we decompose the quantum kinematical system into the Euclidean part and the remaining part (whose phase space admits a compact open subgroup). We then generalize the discrete Hudson theorem \cite{Gro} to the case of totally disconnected 2-regular LCA groups. We also examine angle-number systems with phase space $\tor^n\times\z^n$ and fermionic/hard-core bosonic systems with phase space $\z^{2n}_2$ (which are not 2-regular), and completely characterize their Gaussian states.

\end{abstract}

\maketitle

\tableofcontents

\section{Introduction}
In the phase space formulation of quantum mechanics \cite{Groenewold,Moyal,Weyl,Wigner}, states are represented through Wigner/characteristic functions on the underlying kinematical space, and observables are parametrized by the Weyl representation. Primary examples include systems of $n$-bosonic modes, $n$-qudit systems, and angle-number systems, with associated phase spaces $\Real^{2n}$, $\z_d^{2n}$ and $\tor^n\times\z^n$, respectively. For these systems, phase space methods underlie important concepts and techniques, such as bosonic Gaussian states and channels \cite{wetal}, sharp uncertainty principles \cite{Wer-preprint}, finite-dimensional approximations of continuous systems \cite{DVV,Schwinger}, the stabilizer formalism of quantum error correction \cite{CRSS,Got2}, and the construction of mutually unbiased bases \cite{DEBZ,GHW,Partha04}. Applications of phase space techniques continue to emerge in a variety of systems. In particular, the theory of $p$-adic quantum mechanics \cite{VV} has seen a surge of recent activity in connection with the anti de Sitter/conformal field theory (AdS/CFT) correspondence (see e.g., \cite{BHLL,GKPSW,HMSS}).

Mathematically, quantum kinematical systems with finitely many degrees of freedom are described by a locally compact abelian (LCA) group $G$ and a cocycle $\sigma$. The cocycle induces a symplectic structure on $G$, which encodes the canonical commutation relations of the associated ($\sigma$-projective) Weyl representation. Such abstract quantum kinematical systems have been studied from a variety of perspectives, including finite-dimensional approximations \cite{DHV}, uncertainty relations \cite{Wer16}, and generalized metaplectic operators \cite{Weil}. In this paper we continue this program by developing a formalism to study Gaussian states (and channels) for general quantum kinematical systems. 

Bosonic Gaussian states are defined by the Gaussianity of their associated characteristic functions on the phase space $\Real^{2n}$ (see, e.g., \cite{wetal}). Using the natural notion of Gaussian distribution on LCA groups \cite{Partha1963}, one arrives at a sensible definition of a Gaussian state. However, in many cases of interest (e.g., $G$ finite or totally disconnected), the corresponding class of states is trivial. To overcome this, we advocate the use of Gaussianity in the sense of Bernstein (or B-Gaussianity for short), which is an LCA generalization of Bernstein's classical result: a real
probability distribution $\mu$ is Gaussian if and only if the sum and difference of two independent $\mu$-distributed random variables  are independent \cite{Bernstein}.
Our notion of B-Gaussian states, valid for any phase space $(G,\sigma)$, unifies a variety of examples from the literature, including bosonic Gaussian states, discrete Hudson/stabilizer states \cite{Gro}, vacuum states of $p$-adic oscillator Hamiltonians \cite{VV2}, (classes of) minimal uncertainty states \cite{OP}, and the (relatively) recently introduced Gaussian states for single mode $p$-adic systems \cite{Z,Z2}.

We completely characterize B-Gaussian states over 2-regular (second countable) LCA groups of the form $G=F\times\widehat{F}$ equipped with the canonical normalized 2-cocycle (see Section \ref{sec-first-ex} for the cocycle). Here, 2-regularity means that the doubling map $g\mapsto 2g$ is an automorphism of $G$, and this case includes the systems of $n$-bosonic modes, $n$-qudit systems (for odd $d\geq 3$) and $p$-adic quantum systems.
Thanks to van Kampen's structure theorem, the ``configuration space" $F$ is of the form $\Real^n \times F_c$, where $F_c$ admits a compact open subgroup, and the resulting phase space $G\cong \bR^{2n}\times(F_c\times\widehat{F_c})$. Since the Euclidean case is well understood, we begin by focusing on the case where the phase space is $F_c\times\wh{F_c}$. In this setting, we show that every B-Gaussian state is determined uniquely by a compact open 2-regular isotropic subgroup $H$ of $F_c\times\wh{F_c}$ and a character on $H$ (Theorem \ref{t:finGauss}). We also establish a correspondence between pure B-Gaussian states and symmetric bicharacters on compact open 2-regular subgroups $K$ of $F_c$ (Theorem \ref{thm-finGauss-pure2}), which complements the covariance matrix parametrization in the bosonic setting. As a consequence of our results when $F=F_c$, we show that, amongst B-Gaussian states over general configuration spaces $F=\Real^n \times F_c$, there can be no entanglement across the associated tensor decomposition $L^2(F)=L^2(\bR^{n})\ten L^2(F_c)$ of the system Hilbert space (Theorem \ref{p:noent}). This completes the analysis for 2-regular Weyl systems $G=F\times\widehat{F}$.

In the non-2-regular setting, the structure of B-Gaussian states can be dramatically different. We show that B-Gaussian states over angle-number systems with the phase space $\tor^n\times\z^n$ are forced to be pure, and belong to the canonical ``Fourier'' basis of $L^2(\tor^n)$. Over fermionic and hard-core bosonic systems, which have the same phase space $\z^{2n}_2$ but with different 2-cocycles, we show that there are no B-Gaussian states.

The phase space formulation provides another important function on the phase space for a given quantum state, namely the Wigner function. Wigner functions, which are dual to characteristic functions, are always real-valued and integrate to 1 whenever they are integrable, so they are often called ``pseudo-probability distributions". The natural question of non-negativity of Wigner functions was answered by Hudson for pure states in single-mode bosonic systems \cite{hudson1974wigner}, showing that pure states with non-negative Wigner function are precisely the pure Gaussian states. This was later generalized to multi-mode bosonic systems \cite{soto1983wigner}. Gross continued this line of research, establishing a discrete Hudson's theorem for $n$-qudit systems with odd $d\geq 3$ \cite{Gro}. Our formalism allows one to define Wigner functions in full generality, which, in particular, begs the question of a generalized Hudson's theorem for 2-regular Weyl systems. We partially answer this question by showing that over totally disconnected 2-regular LCA groups of the form $G=F\times\widehat{F}$, a pure state has non-negative continuous Wigner function if and only if it is B-Gaussian.

This paper will be followed by the second part of our project \cite{GQIT2}, which studies Gaussian quantum channels over general kinematical systems and related quantum information theoretic analysis.

\section{Preliminaries on general quantum kinematical systems}\label{sec-gen-Weyl}

\subsection{Locally compact abelian groups}

In this subsection we review the basics of harmonic analysis on locally compact abelian (LCA) groups. All LCA groups in this paper are assumed second countable.

An LCA group $G$ has a dual object called the {\em dual group} $\wh{G}$ consisting of {\em characters} on $G$, i.e., {\em continuous} homomorphisms from $G$ into the {\em circle group} $\tor$. The set $\wh{G}$ is an abelian group with respect to pointwise multiplication, and is locally compact (and second countable) when equipped with the topology of compact convergence. The double dual of an LCA group can be canonically identified with the original group, i.e. we have
    \begin{equation} \notag
        \wh{(\wh{G})} \cong G,
    \end{equation}
    which is known as {\em Pontryagin-van Kampen duality}. Under this duality, properties of $G$ manifest in a dual manner in $\wh{G}$. For instance, an LCA group $G$ is compact if and only if $\wh{G}$ is discrete (\cite[23.17]{HewittRoss}).

For the most part, we use additive notation for LCA groups, so that the group operation will be denoted by $a+b$ for $a,b\in G$ and the identity of $G$ will be denoted by 0. The inverse of $a\in G$ will be denoted by $-a$. However, we will sometimes use multiplicative notation for dual groups $\wh{G}$. For example, the identity element for $\wh{G}$ will be denoted by 1, meaning the constant function with value 1 and the inverse of $\gamma \in \wh{G}$ will be denoted by $\gamma^{-1}$ or $\bar{\gamma}$ (meaning complex conjugate). For $a\in G$ and $\gamma \in \wh{G}$ the duality bracket
\begin{equation} \notag
    \la a, \gamma \ra := \gamma(a) \in \Comp
\end{equation}
will be frequently used. Note that for $\gamma_1, \gamma_2 \in \wh{G}$ and $a_1, a_2\in G$ we have
\begin{equation} \notag
    \la a_1 + a_2, \gamma_1 + \gamma_2 \ra = \la a_1 + a_2, \gamma_1 \ra \la a_1 + a_2, \gamma_2 \ra = \gamma_1(a_1)\gamma_1(a_2)\gamma_2(a_1)\gamma_2(a_2).
\end{equation}

Given a closed subgroup $H$ of $G$ (which we write $H\le G$), the {\em quotient group} $G/H$ is an LCA group endowed with the quotient topology. Its dual group $\wh{G/H}$ can be identified with $H^\perp = \{\gamma \in \wh{G}: \gamma(a) = 1,\;\; a\in H\}$, a closed subgroup of $\wh{G}$ called the {\em annihilator} of $H$.
The identification $H^\perp \cong \wh{G/H}$ (\cite[Theorem 4.39]{Folland-book}) is given by $\gamma \in H^\perp \mapsto \tilde{\gamma}$, where $\tilde{\gamma}(a+H) := \gamma(a)$, $a\in G$. Here, $a+H$ refers to the {\em coset} of $H$ with the representative $a$.
The quotient group $\wh{G}/H^\perp$ can be identified with the dual group $\wh{H}$ through the map $\gamma + H^\perp \in \wh{G}/H^\perp \mapsto \gamma|_H \in \wh{H}$ (\cite[Theorem 4.39]{Folland-book}).
Note that for $H\le G$, the subgroup $H$ is open if and only if $G/H$ is discrete by definition of the quotient topology.

An LCA group $G$ is equipped with a non-zero, translation-invariant Radon measure $\mu = \mu_G$, called the {\em Haar measure}, which is unique up to a positive constant. More precisely, for another non-zero, translation-invariant Radon measure on $G$ we can find $c>0$ such that $\nu = c\cdot \mu$. The choice of Haar measures will be specified later in this paper. When the underlying group $G$ is clear from context, we simply write $\mu$. Otherwise, we use the notation $\mu_G$.

For a closed subgroup $H$ of $G$ the Haar measure provides interesting information about $H$ as follows.
    \begin{equation}\label{eq-cpt-open-Haar}
        \textit{We have $0<\mu_G(H)<\infty$ if and only if $H$ is open and compact.}
    \end{equation}
One direction is trivial by local finiteness of $\mu$ and \cite[Proposition 2.19]{Folland-book}. The converse direction follows from the fact that $\mu|_H$ becomes a finite Haar measure of $H$, so $G/H$ has a $G$-invariant Radon measure $\overline{\mu}$ satisfying $\overline{\mu}(\left \{xH\right\})=\mu(xH)\in (0,\infty)$, which implies discreteness of $G/H$ by \cite[Proposition 1.4.4]{DeEc09}.

The concepts of dual group and Haar measure lead to {\em Fourier transforms}. For $f\in L^1(G):= L^1(G,\mu)$ and $\gamma\in \wh{G}$ we define
    \begin{equation} \notag
     \hat{f}(\gamma) := \int_G f(x)\overline{\gamma(x)}\, d\mu(x),   
    \end{equation}
    and the {\em group Fourier transform} $\F_G$ is defined by
    \begin{equation}\label{eq-group-Fourier}
        \F_G: L^1(G) \to C_0(\wh{G}),\;\; f\mapsto \hat{f},
    \end{equation}
where $C_0(\wh{G})$ refers to the space of all continuous functions on $\wh{G}$ vanishing at infinity. The map $\F_G$ is a norm-decreasing homomorphism with respect to convolution on $L^1(G)$ and pointwise multiplication on $C_0(\wh{G})$, i.e. we have
    \begin{equation} \notag
    \F_G(f*g) = \F_G(f) \cdot \F_G(g),\;\; f,g\in L^1(G),
    \end{equation}
where $f*g$ is the convolution of $f$ and $g$ given by
    \begin{equation} \notag
        f*g(x) = \int_G f(y)g(x-y)d\mu(y),\;\;x\in G.
    \end{equation}
    We will sometimes use the notation $\wh{f}^G$ instead of $\hat{f}$ when we need to specify which group we are referring to. Let us record the special case when $f = 1_K$ for a compact subgroup $K$ of $G$:
    \begin{equation}\label{eq-characteristic}
        \F_G(1_K) = \mu_G(K)1_{K^\perp}.
    \end{equation}
Indeed, we have
    \begin{equation} \notag
        \gamma(y)\int_K \gamma(x) \,d\mu_G(x) = \int_K \gamma(x) \,d\mu_G(x), \forall y\in K
    \end{equation}
so that
    \begin{equation}\label{eq-char-integral}
        \int_K \gamma(x) \,d\mu_G(x) = \begin{cases}\mu_G(K),& \gamma \in K^\perp\\ 0, & \text{otherwise}\end{cases},
    \end{equation}
and this explains \eqref{eq-characteristic}.

The above Fourier transform can be extended to the $L^2(G) = L^2(G,\mu)$-level. More precisely, there is a Haar measure $\mu_{\wh{G}}$ on $\wh{G}$ such that $\F_G: L^1(G)\cap L^2(G) \to L^2(\wh{G})$ is isometric with respect to the corresponding $L^2$-norms. This map can be extended to a unitary (still denoted)
    \begin{equation} \notag
        \F_G: L^2(G, \mu_G) \to L^2(\wh{G}, \mu_{\wh{G}}),
    \end{equation}
    by {\em Plancherel's theorem} (\cite[4.25]{Folland-book}). Note that the choice of $\mu_{\wh{G}}$ depends on $\mu_G$, and we call it the {\em dual Haar measure} to $\mu_G$.

The above $\F_G$ allows for an inverse map at the $L^2$-level, but we have a more direct inversion via the {\em Fourier inversion theorem} (\cite[4.32]{Folland-book}): for $f\in L^1(G)$ such that $\hat{f} \in L^1(\wh{G})$, we have
    \begin{equation}\label{eq-Fourier-inversion}
        f(x) = \int_G \hat{f}(\gamma)\gamma(x)\, d\mu_{\wh{G}}(\gamma),\;\; \text{a.e.}\; x\in G.
    \end{equation}
If, in addition, $f$ is continuous on $G$, then the above identity holds for all $x\in G$. When $f\in L^2(G)$ satisfies $\F_G(f) \in L^1(\wh{G})\cap L^2(\wh{G})$, the function $f$ must be continuous and the above inversion formula also holds by \cite[Theorem 4.4.13]{ReiSte}.

The space $L^1(G)$ embeds naturally into the Banach algebra $M(G)$ of all complex Radon measures on $G$ via the map $f\mapsto f\,d\mu$. The Fourier transform extends to a contraction $\F_G:M(G)\rightarrow C_b(\wh{G})$ satisfying
    \begin{equation} \notag
        \F_G(\nu)(\gamma) = \hat{\nu}(\gamma) := \int_G \overline{\gamma(x)}\, d\mu(x), \ \ \ \nu \in M(G), \ \gamma \in \wh{G},
    \end{equation}
where $C_b(\wh{G})$ is the space of bounded continuous functions on $\wh{G}$. The homomorphism property still holds, i.e. for $\nu_1,\nu_2 \in M(G)$ we have
    \begin{equation}
    \notag     \F_G(\nu_1*\nu_2) = \F_G(\nu_1) \cdot \F_G(\nu_2),
    \end{equation}
    where the convolution $\nu_1*\nu_2$ is determined by the following relation: for any compactly supported continuous function $\phi$ on $G$ we have
    \begin{equation}
     \notag    \int_G \phi \,d(\nu_1*\nu_2) = \int_G\int_G\phi(x)\,d\nu_1(x)d\nu_2(y).
    \end{equation}
We let $M^1(G)$ denote the set of all positive elements in $M(G)$ with total measure 1, namely the {\em (probability) distributions} on $G$. A theorem by Bochner (\cite[33.3]{HewittRoss2}) says that the set $\F_G(M^1(G))$ coincides with the set of all continuous {\em positive definite} functions on $\wh{G}$ having value 1 at the identity. Recall that a function $f: G\to \Comp$ is {\em positive definite} if the matrix $[f(x_i-x_j)]^n_{i,j=1}$ is positive semi-definite for any finite sequence $(x_i)^n_{i=1}\subseteq G$.

The {\em closed support} of $\nu\in M^1(G)$ (which we write $\overline{{\rm supp}}\, \nu$) is defined to be the smallest closed subset $A\subseteq G$ such that $\nu(A) = \nu(G)$. This definition needs to be distinguished with the {\em (open) support} of a continuous function $f$ on $G$, which we write ${\rm supp}\, f$, defined by ${\rm supp}\, f = \{x\in G: f(x) \ne 0\}$. We say that $\nu\in M^1(G)$ is {\em concentrated} on a Borel subset $A\subseteq G$ if $\nu(B) = 0$ for any Borel $B\subseteq G$ such that $A\cap B = \emptyset$.

\begin{prop}\label{prop-posdef}
Let $f: G\to \Comp$ be a continuous positive definite function on an LCA group $G$.
\begin{enumerate}

    \item We have $|f(x)| \le f(0)$ for any $x\in G$.
    
    \item (\cite[Corollary 32.7]{HewittRoss2}) The set $G_1:=\set{x\in G: |f(x)|=f(0)}$ is a closed subgroup of $G$, $|f|$ is constant on the cosets of $G_1$ and $f/f(0)$ is a character on $G_1$. 
\end{enumerate}
\end{prop}

Let us end this subsection by recalling a fundamental structure theorem of LCA groups due to van Kampen: {\it An LCA group $G$ is isomorphic to $\Real^n \times F$ (as topological groups) for some LCA group $F$ containing a compact open subgroup \cite[24.30]{HewittRoss}.}

\subsection{Phase space structure}

Let $G$ be an LCA group equipped with a Borel function $\sigma: G\times G \to \tor$ satisfying the conditions 
    \begin{equation} \notag
    \sigma(a,b)\sigma(a+b,c) = \sigma(a,b+c)\sigma(b,c),\; \sigma(a,0) = \sigma(0,b) = 1,\;\; \mathrm{a.e.} \ a,b,c\in G.    
    \end{equation}
    Note that the above equation holds for almost every $a,b,c\in G$ unless $\sigma$ is continuous. However, we will often omit the expression ``almost every" in the sequel for simplicity. The function $\sigma$ is called a {\em 2-cocycle} (or a {\em multiplier}) on $G$, and determines a  {\em symplectic form} $\Delta:G\times G\to\tor$ via
    \begin{equation}\label{eq-symplectic-form-def}
    \Delta(a,b):= \sigma(a,b)\overline{\sigma(b,a)},\; a,b\in G.
    \end{equation}
Note that $\Delta$ is a {\em bicharacter}, meaning that $\Delta$ is continuous and $\Delta(\cdot,b)$ and $\Delta(a,\cdot)$ are characters on $G$ for all $a, b\in G$ \cite[p.533]{DigernesVaradarajan04}. Note that Borel measurability of $\sigma$ and $\Delta$ being Borel homomorphism in each argument guarantees that $\Delta$ is continuous \cite[p.281]{Mackey}. We require the map $\Phi_{\Delta}: G \to \widehat{G}$ given by
    \begin{equation}\label{eq-symplectic-iso}\Phi_{\Delta}(a)(b) = \Delta(a,b),\; a,b\in G
    \end{equation}
to be a topological group isomorphism, in which case we call the associated 2-cocycle $\sigma$ a {\em Heisenberg multiplier} (following the terminology of \cite{DigernesVaradarajan04}). The pair $(G,\sigma)$ (or rather $(G,\Delta)$) is viewed as the {\em phase space} underlying a {\em general quantum kinematical system} (see, e.g., \cite{DigernesVaradarajan04}).

For example, the standard choice of 2-cocycle on the system of $n$-bosonic modes with the phase space $G = \mathbb R^{2n} \cong \Real^n\times \Real^n$ is given by
    \begin{equation}\label{eq-Euclid-2-cocycle}
    \sigma_{\rm boson}( a ,  b ) =  \exp\left(-\frac{i}{2} a ^T J  b \right), \;\; a ,  b  \in G, 
    \end{equation}
where $J={\footnotesize \begin{bmatrix}0 & I_n\\ -I_n & 0\end{bmatrix}}\in M_{2n}(\Real)$ is the matrix of the canonical symplectic form on $\Real^{2n}$. 
Note also that the above map $\Phi_\Delta$ is different from the usual identification $x \in \Real^{2n}\mapsto \gamma_x \in \widehat{\Real^{2n}}$ given by $\gamma_x(y):=e^{i \la x, y\ra}$, $y \in \Real^{2n}$, %(\tcr{do we want $2\pi$ in the argument of $e$?}),
which we call the {\em canonical identification}.

From the fact that $\Phi_\Delta(a)(a) = 1$ for any $a\in G$, the isomorphism $\Phi_\Delta$ is called a {\em symplectic self-duality} for $G$ \cite{PrasadShapiroVemuri10}. A typical example of an LCA group $G$ with symplectic self-duality is $G=F\times \widehat{F}$ for another LCA group $F$, and this is exactly the class we will focus on. Note, however, that there exist LCA groups with symplectic self-duality not isomorphic to $F\times \widehat{F}$ for any LCA group $F$ \cite[Theorem 11.2]{PrasadShapiroVemuri10}.

Since $\sigma$ is a Heisenberg multiplier, there is a unique (up to unitary equivalence) irreducible unitary projective representation with respect to $\sigma$ (or $\sigma$-representation) $W: G\to \mathcal{U}(\Hi_W)$ for some Hilbert space $\Hi_W$ \cite[Theorem 2]{DigernesVaradarajan04}. Being a $\sigma$-representation means that the map $a\in G \mapsto W(a)\psi$ is Borel for any $\psi\in \Hi_W$ and we have
    \begin{equation}\label{eq-proj-rep}
    W(a)W(b) = \sigma(a,b)W(a+b),\; a,b \in G.
    \end{equation}
Note that there are important examples of discontinuous 2-cocycles as we can see in Section \ref{ss:ANS}.

We call $W$ and $W(a)$, $a\in G$, the {\em Weyl representation} and the {\em Weyl operators} following the standard terminology. Note that the Weyl operators satisfy the {\em canonical commutation relations (CCR)}
    \begin{equation} \label{eq-Weyl-CCR}
        W(a) W(b) =  \Delta(a,b)W(b)W(a),\;\;  a ,  b \in G.
    \end{equation}
See \cite[\S 3.5]{DigernesVaradarajan04} for a concrete model of $\Hi_W$ and $W$.

A 2-cocycle $\sigma$ on an LCA group $G$ is {\em normalized} if $\sigma(a,-a)=1$, $a\in G$. This additional requirement on $\sigma$ is essential to accommodate ``Gaussian states" as we can see in Remark \ref{rem-normalized-is-needed}(4) below.
Fortunately, any 2-cocycle $\sigma$ allows a {\em normalization} $\tilde{\sigma}$, which is similar to $\sigma$ as 2-cocycles in the sense that there exists a Borel function $\xi:G\to \tor$ (called a {\em normalizing factor}) so that
    \begin{equation}\label{eq-equivalent-cocycle}
        \tilde{\sigma}(a,b)=\frac{\xi(a)\xi(b)}{\xi(a+b)}\sigma(a,b),\;\; a,b\in G,
    \end{equation}
defines a normalized 2-cocycle.
In this case, the 2-cocycles $\sigma$ and $\tilde{\sigma}$ determine the same symplectic form $\Delta(a,b)=\sigma(a,b)\overline{\sigma(b,a)} = \tilde{\sigma}(a,b)\overline{\tilde{\sigma}(b,a)}$, and therefore $\sigma$ is a Heisenberg multiplier if and only if $\tilde{\sigma}$ is. Moreover, if $W$ is an irreducible $\sigma$-representation of $G$ acting on $\Hi_W$, then \begin{equation}
  \notag   W_{1/2}(a):=\xi(a)W(a)
\end{equation}
is an irreducible $\tilde{\sigma}$-representation of $G$ acting on the same Hilbert space $\Hi_{W}$. We will take $\xi$ to be a Borel measurable {\em square root} of the function $a\in G \mapsto \overline{\sigma(a,-a)}$, hence the $1/2$ in the notation $W_{1/2}$.
Note that a choice of square root is always possible but not unique, in general. Thus, the choice of $\xi$ and $\tilde{\sigma}$ will be specified whenever necessary.

\subsection{Weyl systems}\label{sec-first-ex}

The main class of quantum kinematical systems we consider have the form $G=F\times \widehat{F}$ for an LCA group $F$. Such groups admit a {\em canonical} choice of 2-cocycle, $\sigma_{\rm can}: G\times G \to \tor$ given by
    \begin{equation}\label{eq-can-2-cocycle}
        \sigma_{\rm can}((x,\gamma), (x', \gamma')) :=  \gamma(x'),\; x,x'\in F,\; \gamma, \gamma'\in \widehat{F}.
    \end{equation}
It is straightforward to see that $\sigma_{\rm can}$ is a Heisenberg multiplier and we call the pair $(F\times \widehat{F}, \sigma_{\rm can})$ a {\em Weyl system}. The group $F$ is called the {\em configuration space}.

In this case we have a simple description for the unique irreducible $\sigma_{\rm can}$-representation $W = W_{\rm can}$ as follows \cite{Prasad11}. We first define the {\em translation operator} $T_x$ and the {\em modulation operator} $M_\gamma$ for $x\in F$ and $\gamma \in \widehat{F}$ acting on $\Hi_W := L^2(F)$ by
	\begin{equation}
\notag 	    T_xf(y) := f(y-x),\;\; M_\gamma f(y) := \gamma(y)f(y),\; f\in L^2(F),\; y\in F.
	\end{equation}
	Then, $W:G \to \mc B(L^2(F))$ is given by
	\begin{equation}
	\notag     W(x,\gamma) := T_xM_\gamma,\; (x,\gamma)\in G.
	\end{equation}
%(\tcr{Most authors that I am aware of (e.g., Folland, Holevo, Weil) use $M_\gamma T_x$. The resulting $W$ would be projective with respect to the 1,4-paring on $G\times G$, as opposed to the 2,3 pairing used above. I'm fine with the above choice, so long as we are consistent and the correct formulas are used})

\subsubsection{2-Regular groups}
The above 2-cocycle $\sigma_{\rm can}$ is never normalized unless $G$ is trivial. There is a canonical normalization when the group $G=F\times\widehat{F}$ (equivalently, $F$) is {\it 2-regular}. Here, we say that the abelian group $G$ is 2-regular if the map $a  \mapsto 2 a $ is an automorphism of $G$, and we denote its inverse by $2^{-1}$. In this case, there is a unique bicharacter $\xi$ such that $\xi(x,\gamma)^2=\la x,\gamma\ra$ \cite[Lemma 1]{DigernesVaradarajan04}, namely
    \begin{equation}
  \notag   \xi(x,\gamma):=\la x,\gamma \ra ^{1/2}:=\la 2^{-1}x,2^{-1}\gamma \ra^2=\la x,2^{-1}\gamma \ra=\la 2^{-1}x,\gamma \ra.    
    \end{equation}
    With this $\xi$ as the normalization factor, we get the
{\em canonical normalization} $\tilde{\sigma}_{\rm can}$ of $\sigma_{\rm can}$ given by
\begin{equation}\label{eq-normalization}
	\tilde{\sigma}_{\rm can}( a , b ) :=  \Delta(2^{-1} a ,2^{-1} b )^2 =\Delta( a ,2^{-1} b ) =\Delta(2^{-1} a ,b ), \; a,b\in G.
\end{equation}
We sometimes write $\tilde{\sigma}_{\rm can} = \Delta^{1/2}$ for an obvious reason, and we also call the pair $(F\times \widehat{F}, \tilde{\sigma}_{\rm can})$ a {\em Weyl system}.
Note finally that the corresponding Weyl representation $W_{1/2}$ becomes
    \begin{equation}
  \notag  W_{1/2}(x,\gamma)\psi(y)=\la x,\gamma \ra^{1/2} T_x M_{\gamma}\psi(y)=\la x,\gamma \ra^{-1/2} \la y, \gamma\ra \psi(y-x),\;\; \psi\in L^2(F).
    \end{equation}
    
\begin{ex} (Bosonic systems) The additive group $\Real^n$ is 2-regular, and if we identify $\widehat{\Real^n}\cong {\Real^n}$ via $\la x, \gamma_y\ra:=e^{i \la x, y\ra},\; x,y\in \Real^n$, then the formula \eqref{eq-Euclid-2-cocycle} is recovered with $\sigma_{\rm boson}=\tilde{\sigma}_{\rm can}$. The corresponding symplectic form satisfies
\begin{equation}
\notag     \Delta(z,z')=e^{i \la y, x'\ra}e^{-i\la y',x\ra}=e^{i\la Jz,z'\ra}, \ \ \ z=(x,y), \ z'=(x',y')\in\Real^{2n},
\end{equation}
where $J={\footnotesize \begin{bmatrix}0 & I_n\\ -I_n & 0\end{bmatrix}}\in M_{2n}(\Real)$
and $\la \cdot, \cdot \ra$ refers to the usual inner product on Euclidean spaces. The Weyl representation becomes
\begin{equation}
\notag     W_{1/2}(x,y)\psi(t)=e^{-\frac{i}{2} \la x, y\ra}e^{i \la y, t\ra} \psi(t-x),\;\; \psi\in L^2(\Real^n), \ x,y\in\Real^n.
\end{equation}
This is equivalent to the Weyl representation used in \cite[\S 12.2]{Hol} and \cite[\S 1.3]{Folland-book}, for example.
\end{ex}

\begin{ex}\label{ex:qudit} (Qudit systems) If $d\geq 3$ is an odd integer then $\z_d^n$ is a finite 2-regular abelian group. ($2^{-1}=\frac{d+1}{2}$ is the multiplicative inverse of 2 in the ring $\z_d$.) Similar to above, we have the self-duality $\widehat{\z_d^n}\cong\z_d^n$ via
\begin{equation}\label{e:char1}\gamma_y(x)=e^{\frac{2\pi i}{d}\la x,y\ra}, \ \ \ x,y\in\z_d^n.\end{equation}
Under the canonical identification 
\begin{equation}
\notag     \ell^2(\z_d^n)=\ell^2(\z_d)\ten\cdots\ten\ell^2(\z_d)\cong\bC^d\ten\cdots\ten\bC^d,
\end{equation}
the corresponding multiplication operators $M_y:=M_{\gamma_y}$ satisfy
\begin{equation}
\notag M_y=Z^{y_1}\ten\cdots\ten Z^{y_n}, \ \ \ y=(y_1,...,y_n)\in\z_d^n,
\end{equation}
where $Z:\bC^d\ni |k\ra\mapsto e^{\frac{2\pi ik}{d}}|k\ra\in\bC^d$ is the qudit generalization of the Pauli $Z$ matrix.

Similarly, the translation operators are given by
\begin{equation}
\notag T_x=X^{x_1}\ten\cdots \ten X^{x_n}, \ \ \ x=(x_1,...,x_n)\in\z_d^n,    
\end{equation}
where $X:\bC^d\ni |k\ra\mapsto |k+1\ra\in\bC^d$ is the qudit generalization of the Pauli $X$ matrix. The Weyl representation $W:\z_d^n\times\z_d^n\rightarrow\mc{B}((\bC^d)^{\ten n})$ is then simply
\begin{equation}
 \notag    W(x,y)=e^{\frac{(d+1)\pi i}{d}\la x,y\ra}X^{x_1}Z^{y_1} \ten\cdots \ten X^{x_n}Z^{y_n}, \ \ \ x,y\in\z_d^n.
\end{equation}
In this case the symplectic form satisfies
\begin{equation}
\Delta((x,y),(x',y'))=e^{\frac{2\pi i}{d}(\la y,x'\ra-\la y',x\ra)}, \ \ \ x,y,x',y'\in\z_d^n.    
\end{equation}
\end{ex}
 
\begin{ex}\label{ex:p-adic} ($p$-adic systems) If $p$ is a prime, the field of $p$-adic numbers $\textbf{Q}_p$ is a 2-regular totally disconnected abelian group, along with any finite product $\textbf{Q}_p^n$. It is well-known that $\widehat{\textbf{Q}_p}\cong\textbf{Q}_p$ via the duality
\begin{equation}
 \notag    \la x,y\ra=e^{2\pi i\{xy\}_p}, \ \ \ x,y\in\textbf{Q}_p,
\end{equation}
where $\{x\}_p$ is the fractional part of $x$ defined through the (unique) power series representation of $x$ as follows:
\begin{equation}
\notag \{x\}_p=\sum_{n=-k}^{-1}x_n p^n, \ \ \textnormal{when} \ \ x=\sum_{n=-k}^\infty x_np^n.    
\end{equation}
(see \cite[Theorem 4.12]{Folland-book}, for instance). The symplectic structure on $G=\textbf{Q}_p^n\times\textbf{Q}_p^n$ is given similarly as 
\begin{equation}
\notag \Delta((x,y),(x',y'))=\prod_{k=1}^ne^{2\pi i(\{y_kx_k'\}_p-\{y_k'x_k\}_p)}, \ \ \ x,y,x',y'\in \textbf{Q}_p^n.    
\end{equation}
\end{ex}

\subsubsection{Weyl systems over non-2-regular groups I: Angle-number systems}\label{ss:ANS}
When the LCA group $G=F\times\widehat{F}$ is not 2-regular the canonical normalization \eqref{eq-normalization} is no longer available. Instead, we will specify a normalization $\tilde{\sigma}_{\rm can}$ of $\sigma_{\rm can}$ for each individual case.

%\tcr{(Is it desirable to change $\tor^d\times \z^d\to \tor^n\times \z^n$? we are already using the letter $n\in \z^d$...)} 

We call the quantum system described by $(\tor^d \times \z^d, \tilde{\sigma}_{\rm can})$ the {\em angle-number system in $d$-modes}, which we named after \cite[Table I]{Wer16}. Note that there are many physical quantum systems modelled through the angle-number system in $1$-mode such as the quantum rotor \cite{RKSE} and the dynamics of a Josephson junction between two isolated islands \cite{G}. The case $d=2$ for two rotors can be found in \cite[Sec IV. B.]{ACP}.

The canonical 2-cocycle becomes
    $$\sigma_{\rm can}((\theta, n), (\theta',n')) = e^{2\pi i \la  \theta', n \ra},\;\; (\theta, n)\in \tor^d \times \z^d.$$
Here, we identify $\tor \cong \left[-\frac{1}{2},\frac{1}{2}\right)$ and for $\theta = (\theta_1,\cdots, \theta_d) \in \tor^d \cong \left[-\frac{1}{2},\frac{1}{2}\right)^d$ and $n = (n_1,\cdots, n_d) \in \z^d$ we have
\begin{equation} \label{eq-angle-number-duality}
    \la  \theta, n \ra := n_1\theta_1 + \cdots + n_d\theta_d \in \Real.
\end{equation}
Our choice of normalizing factor $\xi$ is
    \begin{equation}\label{eq-xi-choice}
        \xi(\theta,n)=e^{\pi i\la \theta, n\ra},\; (\theta,n)\in \tor^d\times \z^d \cong \left[-\frac{1}{2},\frac{1}{2}\right)^d \times \z^d.
    \end{equation}
Some care needs to be applied here since the identification $\tor \cong \left[-\frac{1}{2},\frac{1}{2}\right)$ does not respect the group structure of $\tor$ and $\xi$ is discontinuous at $(\theta, n)$ when $\theta_j = -\frac{1}{2}$ for some $1\le j\le d$, so that the resulting normalization $\tilde{\sigma}_{\rm can}$ is also discontinuous there.

In this case the associated Weyl representation $W_{1/2}$ becomes
    $$W_{1/2}(\theta,n):=e^{\pi i\la  \theta, n \ra}T_\theta M_n,\;\; (\theta,n)\in \tor^d\times \z^d,$$
which are operators acting on the Hilbert space $\Hi = L^2(\tor^d)\cong \ell^2(\z^d)$ with the canonical choice of orthonormal basis $\{|e_m\ra : m\in \z^d\}$, where $e_m(\theta) = e^{2\pi i \la \theta, m\ra}$, $\theta \in \tor^d$. We will simply write $|m\ra$ for $|e_m\ra$.

\subsection{Fermions and hardcore bosons}\label{ss:fermion}

In this section we examine two quantum kinematic systems over the phase space $G=\z_2^n\times \widehat{\z_2^n}\cong \z_2^{n}\times \z_2^n=\z_2^{2n}$.

\subsubsection{Fermionic systems} \label{sec-fermions}
Even though our phase space is of the form $F\times \widehat{F}$, we can endow a 2-cocycle which is not similar to the canonical one (when $n\ge 2$). More precisely, our choice of 2-cocycle is as follows.
    \begin{equation}\label{eq-fer-2-cocycle}
       \sigma_{\rm fer}(a,b) := (-1)^{a^T A b},\; a,b \in \z^{2n}_2,
   \end{equation}
where $A={\Tiny \begin{bmatrix} 0 &&&\\ 1& 0&&\\1&1&0&\\ \vdots& \vdots& \ddots & \ddots & \\1 & 1 & \cdots & 1 & 0 \end{bmatrix}}.$ Note that the 1-mode case (i.e. $n=1$) goes back to the canonical 2-cocycle on $\z_2\times \z_2$. We can check that $\sigma_{\rm fer}$ is a Heisenberg multiplier by observing that $A+A^T$ is invertible. Indeed, we have $A+A^T={\Tiny \begin{bmatrix} \Om &E &E &\cdots &E \\ E &\Om &E &\cdots &E \\ E &E &\Om &\cdots & E \\ \vdots &\vdots &\vdots & \ddots &\vdots \\ E &E &E & \cdots &\Om \end{bmatrix}}$, where $E={\Tiny \begin{bmatrix} 1&1\\ 1&1 \end{bmatrix}}$, $\Om={\Tiny \begin{bmatrix} 0&1\\ 1&0 \end{bmatrix}}$. Then the invertibility of $A+A^T$ is direct from the matrix identity $(I+A)(A+A^T)(I+A^T)=\bigoplus^n_{j=1}\Om$, where we use the relations $\Om E=E\Om=E$ and $E^2=2E=0$.

The quantum kinematical system $(\z_2^n\times \widehat{\z_2^n}, \sigma_{\rm fer})$ describes a {\it fermionic system in $n$-modes}. For a detailed explanation, let us recall the {\em Majorana operators} $c_1 , \dots, c_{2n}$, which are self-adjoint operators acting on $\Hi = \Comp^{2^n}= \ell^2(\z^n_2)$ satisfying the CAR (canonical anti-commutation relations): 
    $$\{c_j, c_k\} = 2\delta_{jk},\; 1\le j,k\le 2n.$$
Note that $c_j$'s are realized as
\begin{align*}
    c_{2j-1}=Y\otimes\cdots\otimes Y\otimes X\otimes I\otimes \cdots \otimes I\\
    c_{2j}=Y\otimes\cdots\otimes Y\otimes Z\otimes I\otimes \cdots \otimes I,
\end{align*}
where $X$ and $Z$ appear at the $j$-th tensor component and $X$, $Y$, $Z$ are the $2\times 2$ \textit{Pauli matrices}.
The unique irreducible unitary $\sigma_{\rm fer}$-representation $W_{\rm fer}:\z^{2n}_2 \to \mc{U}(2^n)$ is given by
    \begin{equation}\label{eq-W-eps}
    W_{\rm fer}( a ) := c^{x_1}_1 \cdots c^{x_{2n}}_{2n},\;\;  a  = (x_1,\cdots,x_{2n}) \in \z^{2n}_2.    
    \end{equation}
That $W_{\rm fer}$ is a $\sigma_{\rm fer}$-representation is straightforward to check. Irreducibility follows from the fact that $\{W_{\rm fer}( a ):  a  \in \z^{2n}_2\}$ forms an orthogonal basis of $M_{2^n}(\Comp)$ with respect to the trace inner product. 
%For the irreducibility of $W_{\rm fer}$ we may appeal to the twisted version of Schur's lemma as follows: let $X$ be an intertwiner of $W_{\rm fer}$, then we know that $X$ commutes with each of $W_{\rm fer}( a )$, $ a  \in \z^{2n}_2$ and consequently with span$\{W_{\rm fer}( a ):  a  \in \z^{2n}_2\} = M_{2^n}(\Comp)$, so that it is a scalar multiple of the identity. The latter equality comes from the fact that $\{W_{\rm fer}( a ):  a  \in \z^{2n}_2\}$ forms an orthogonal basis of $M_{2^n}(\Comp)$ with respect to the trace inner product.

We  consider a normalization $\tilde{\sigma}_{\rm fer}$ of $\sigma_{\rm fer}$ given by
    \begin{equation}\label{eq-normalization-fermi}
        \tilde{\sigma}_{\rm fer}( a , b ):=\frac{\xi( a )\xi( b )}{\xi( a + b )}\sigma_{\rm fer}( a , b ),\;\; a,b \in \z^{2n}_2,
    \end{equation}
where the normalizing factor $\xi:\z_2^{2d}\to \tor$ is chosen to satisfy
    $$\xi( a )^2=\xi(a)\xi(-a)=\tilde{\sigma}_{\rm fer}( a ,- a )\overline{\sigma_{\rm fer}( a ,- a )}=(-1)^{ a ^T A a },\;\; a\in \z^{2n}_2.$$
Note that there are many choices for the factor $\xi$. We will not fix a particular choice of $\xi$ for fermionic systems, but instead consider all possible choices of $\xi$ (see Section \ref{sec-second-ex}).

Finally, we remark that the unique irreducible $\tilde{\sigma}_{\rm fer}$-representation is
    $$W_{1/2,{\rm fer}}:=\xi W_{\rm fer}.$$

\subsubsection{Hardcore bosons}\label{subsubsec-hard-core-boson}
Here we consider the canonical 2-cocycle $\sigma_{\rm can}$ \eqref{eq-can-2-cocycle} on $G = \z^n_2 \times \widehat{\z^n_2} \cong \z^{2n}_2$. As in the qudit case (Example \ref{ex:qudit}), the associated Weyl operators have the following form:
\begin{equation}\label{eq-W-hardcore}
    W_{\rm can}(x,y)= X^{x_1}Z^{y_1}\otimes \cdots \otimes X^{x_n}Z^{y_n} = h^{x_1}_1h^{y_1}_2h^{x_2}_3h^{y_2}_4\cdots h^{x_n}_{2n-1}h^{y_n}_{2n},
\end{equation}
where $X$, $Z$ are the standard $2\times 2$ Pauli matrices and the matrices $h_j$, $1\le j\le 2n$ are given by
\begin{align*}
    h_{2j-1} = I\otimes \cdots \otimes I \otimes X \otimes I \otimes \cdots \otimes I\\
    h_{2j} = I\otimes \cdots \otimes I \otimes Z\otimes I\otimes \cdots \otimes I,
\end{align*}
where $X$ and $Z$ appear at the $j$-th tensor component for $1\le j\le n$.
The self-adjoint matrices $h_j$, $1\le j\le 2n$ are an analogue of Majorana operators and they satisfy
    $$h_kh_l = - h_lh_k,\;\; (k,l) = (2j-1,2j)\; \text{or}\; (2j,2j-1),\; 1\le j\le n$$
and $h_kh_l = h_lh_k$ for other choices $(k,l)$. In other words, the observables $h_j$, $1\le j\le 2n$ anti-commute in the same modes and commute for different modes, and the associated quantum system corresponds to {\em ``hardcore bosons"} of $n$ degrees of freedom \cite[Section II]{CL}.

To apply our program in this setting, we consider a normalization $\tilde{\sigma}_{\rm can}$ of $\sigma_{\rm can}$ given by
    \begin{equation}\label{eq-normalization-hardcore}
        \tilde{\sigma}_{\rm can}( a , b ):=\frac{\xi( a )\xi( b )}{\xi( a + b )}\sigma_{\rm can}( a , b ),\;\; a,b \in \z^{2n}_2,
    \end{equation}
where the normalizing factor $\xi:\z_2^{2n}\to \tor$ is chosen to satisfy
    $$\xi( a )^2=\xi(a)\xi(-a)=\tilde{\sigma}_{\rm can}( a ,- a )\overline{\sigma_{\rm can}( a ,- a )}=(-1)^{ a ^T L a },\;\; a\in \z^{2n}_2,\;\; L:=\tiny\begin{bmatrix} 0 & 0 \\ I_n & 0 \end{bmatrix}.$$
As in the fermionic system, we will not fix a particular choice for $\xi$, and the unique irreducible $\tilde{\sigma}_{\rm can}$-representation is given by
$W_{1/2,{\rm can}}:=\xi W_{\rm can}$.

\section{Characteristic and Wigner functions of quantum states}

Throughout this and the next section we fix a general quantum kinematical system given by the pair $(G,\sigma)$ consisting of a second countable LCA group $G$ and a normalized 2-cocycle $\sigma$ which is a Heisenberg multiplier.

Similar to the bosonic case (e.g. \cite[\S 12]{Hol}) and certain qudit systems (e.g. \cite{Gro}), quantum states on $\Hi := \Hi_W$ -- the irreducible representation space of $W$ -- can be recovered through their {\em characteristic functions} on the phase space $G$.

Recall that the set of all quantum states on $\Hi$ (denoted by $\D = \D(\Hi)$) is a subset of $\mc S^1(\Hi)$, the \textit{trace class} on $\Hi$ equipped with the trace norm $\|X\|_1 = {\rm Tr}(|X|) ={\rm Tr}((X^*X)^{\frac{1}{2}})$, $X\in \mc S^1(\Hi)$. Note that $\mc S^1(\Hi)$ is a subspace of $\mc S^2(\Hi)$, the {\it Hilbert-Schmidt class} on $\Hi$ equipped with the Hilbert-Schmidt norm $\|X\|_2 = ({\rm Tr}(X^*X))^{\frac{1}{2}}$, $X\in \mc S^2(\Hi)$.

	\begin{defn} \label{defn-chftn}
		Let $\rho \in \mc S^1(\Hi)$. Its {\em characteristic function} $\chi_\rho \in L^\infty(G)$ is defined by
			$$\chi_\rho( a ) = {\rm Tr}(W_{1/2}( a )^*\rho),\;\; a\in G.$$
		For a pure state $\rho = |\psi \ra \la \psi |$ with $\psi \in \Hi$, we will simply write $\chi_\psi$ instead of $\chi_{|\psi \ra \la \psi |}$.	
	\end{defn}

It is straightforward that $\|\chi_{\rho}\|_{\infty}\leq \|\rho\|_1$, so $\chi_\rho$ is indeed bounded. The terminology ``characteristic function'' can be justified from the fact that $\chi_\rho$ determines the original operator $\rho$ via the {\em twisted group Fourier transform} on $G$. See \cite{KleppnerLipsman72} and \cite{Mackey} for details of twisted group Fourier transforms on locally compact (not necessarily abelian) groups. In our specific situation, namely that $W$ is the only (up to unitary equivalence) $\sigma$-representation, the theory simplifies.

	\begin{defn}
	The \textit{twisted group Fourier transform} $\F^\sigma_G$ on $G$ is given by
	    \begin{equation}\label{eq-twisted-Fourier-transform}
	        \F^\sigma_G: L^1(G) \to \B(\Hi_\weyl),\;\; f\mapsto \hat{f}(W_{1/2}):= \int_G f(a) W_{1/2}(a) d\mu(a) \in B(\Hi_W),
	    \end{equation}
where the choice of Haar measure $\mu$ on $G$ will be specified below in Theorrem \ref{thm-twistedPlancherel}.
	\end{defn}

The map $\F^\sigma_G$ is a norm-decreasing $*$-homomorphism with respect to {\em twisted convolution} and {\em twisted involution}, defined respectively by
    $$(f *_\sigma g)(a) := \int_G f(b) g(a-b) \sigma(b,a-b) d\mu(b),\; a\in G,$$
and
    $$f^{\star\sigma}(a) := \overline{\sigma(a,-a) f(-a)},\; a\in G,$$
for $f,g \in L^1(G)$. More precisely, we have $\F^\sigma_G(f *_\sigma g) = \F^\sigma_G(f)\cdot \F^\sigma_G(g)$ as the product (or composition) of two operators and $\F^\sigma_G(f^{\star\sigma}) = \F^\sigma_G(f)^*$ as the adjoint operator for $f,g \in L^1(G)$. It extends to a unitary operator acting on $L^2(G)$.

\begin{thm}\label{thm-twistedPlancherel} ({\bf Twisted Plancherel theorem}, \cite[Theorem 7.1]{KleppnerLipsman72})
The twisted group Fourier transform $\mc F^\sigma_G$ extends to a unitary equivalence between $L^2(G)$ and $\mc S^2(\Hi_W)$ for a suitable choice of Haar measure $\mu$ on $G$. In particular, we have
    \begin{equation}\label{eq-twisted-Plancherel}
        \int_G f\bar{g} d\mu = {\rm Tr}(\hat{f}(W_{1/2}) \hat{g}(W_{1/2})^*), \;\; f,g\in L^1(G)\cap L^2(G).
    \end{equation}
Moreover, the extended map $\mc{F}_\sigma$ intertwines the {\em left regular $\sigma$-representation} $\lambda_\sigma: G \to \mc B(L^2(G))$ given by
    \begin{equation*}\label{eq-reg-rep}
        \lambda_\sigma(a)f(b) = \sigma(a,b-a)f(b-a),\; a,b\in G,\; f\in L^2(G),
    \end{equation*}
with an amplification of $W_{1/2}$. More precisely, we have
\begin{equation}\label{eq-intertwining}
[\F^\sigma_G \circ \lambda_\sigma(a)](f) = W_{1/2}(a)\cdot [\F^\sigma_G(f)]:L^2(G)\rightarrow \mc S^2(\mathcal{H}_W),\; a\in G,\; f\in L^2(G).
\end{equation}
\end{thm}

In what follows, we fix the Haar measure $\mu$ on $G$ respecting \eqref{eq-twisted-Plancherel}.

The following \textit{twisted Fourier inversion} justifies the ``characteristic function''  terminology, and will be useful in Section \ref{sec-2-regular}.

\begin{prop}\label{prop-Fourier-char}
For any $\rho \in \mc S^1(\Hi)$ we have $\chi_\rho \in L^2(G)$ and $\F^\sigma_G(\chi_\rho) = \rho.$
\end{prop}
\begin{proof}
Since $\F^\sigma_G: L^2(G) \to \mc S^2(\Hi)$ is unitary, span$\{\F^\sigma_G(\varphi): \varphi \in C_c(G)\}$ is dense in $S^2(\Hi)$, where $C_c(G)$ is the space of all continuous functions on $G$ whose closed support is compact. Consequently, span$\{\F^\sigma_G(\varphi_1)\F^\sigma_G(\varphi_2)^*: \varphi_1, \varphi_2 \in C_c(G)\}$ is dense in $S^1(\Hi)$. 

First, for $\rho = \F^\sigma_G(\varphi_1)\F^\sigma_G(\varphi_2)^* = \F^\sigma_G(\varphi_1 *_\sigma \varphi^{\star_\sigma}_2)$ with $\varphi_1, \varphi_2 \in C_c(G)$, the intertwining relation \eqref{eq-intertwining} with $\lm_\sigma$ entails
    \begin{equation}\label{eq30}
        \varphi_1 *_\sigma \varphi^{\star \sigma}_2(\cdot) = \la \varphi_1 | \lambda_\sigma(\cdot)\varphi_2\ra = \textrm{Tr}(W_{1/2}(\cdot)^*\F^\sigma_G(\varphi_1)\F^\sigma_G(\varphi_2)^*) = \chi_\rho(\cdot).
    \end{equation}

For arbitrary $\rho \in \mc S^1(\Hi)$, there exist a sequence $(\rho_n)_n$ in the space 
    $$\text{span}\{\F^\sigma_G(\varphi_1)\F^\sigma_G(\varphi_2)^*: \varphi_1, \varphi_2 \in C_c(G)\}$$
such that $\displaystyle \lim_{n\rightarrow \infty}\|\rho-\rho_n\|_{S^1(\Hi)}=0$. Since $\chi_{\rho_n}=(\F^\sigma_G)^{-1}(\rho_n)$ from \eqref{eq30}, we have
\begin{align*}
    \lim_{n\rightarrow \infty}\|(\F^\sigma_G)^{-1}(\rho)-\chi_{\rho_n}\|_{L^2(G)}&=\lim_{n\rightarrow \infty}\|(\F^\sigma_G)^{-1}(\rho)-(\F^\sigma_G)^{-1}(\rho_n)\|_{L^2(G)}\\
    & \leq \lim_{n\rightarrow \infty}\|\rho-\rho_n\|_{S^1(\Hi)}=0.
\end{align*}
In particular, the $L^2$-convergence of $(\chi_{\rho_n})_n$ to $(\F^\sigma_G)^{-1}(\rho)$ implies that a subsequence of $(\chi_{\rho_n})_n$ converges to $(\F^\sigma_G)^{-1}(\rho)$ almost everywhere. On the other hand, the condition $\displaystyle \lim_{n\rightarrow \infty}\|\rho-\rho_n\|_{S^1(\Hi)}=0$ implies $\displaystyle \lim_{n\rightarrow \infty}\|\chi_{\rho_n}-\chi_\rho\|_{\infty}=0$. Thus, $\chi_{\rho}=(\F^\sigma_G)^{-1}(\rho)\in L^2(G)$. 
\end{proof}

\begin{rem}\label{rem-typeI-condition}${}$
    \begin{enumerate}
    
         \item For $f\in L^2(G)$, the element $\F^\sigma_G(f)$ is originally defined by the $\mc S^2(\Hi)$-limit of $\F^\sigma_G(f_n) = \hat{f}_n(W_{1/2})$ for some sequence $(f_n) \subseteq L^1(G)\cap L^2(G)$ converging to $f$ in $L^2(G)$. However, we may still express the element $\F^\sigma_G(f)$ via the integral representation $\int_G f( a ) W_{1/2}( a ) d\mu( a )$ once we understand it as a bounded operator on $\Hi$ given in the weak sense.
         Indeed, for any $\xi, \eta \in \Hi$ we have $\|\chi_{|\xi\ra\la \eta|}\|_{L^2(G)} = \| |\xi\ra\la \eta| \|_{\mc S^2(\Hi)} = \|\xi\| \cdot \|\eta\|$ by Proposition \ref{prop-Fourier-char}.
         Thus,
            \begin{align*}
            |\la\eta |\int_{G}f( a )W_{1/2}( a )d\mu( a )|\xi\ra|
            & = \left |\int_{G}f( a )\la\eta |W_{1/2}( a )|\xi\ra d\mu( a )\right |\\
            & = \left |\int_{G}f( a )\chi_{|\xi\ra\la \eta|}(a) d\mu( a )\right |\\
            & \le \|f\|_{L^2(G)} \|\xi\| \cdot \|\eta\|.
            \end{align*}
        This explains that the integral $\int_G f(a) W_{1/2}(a) d\mu(a)$ defines a bounded operator on $\Hi$ in the weak sense. The same computation also tells us that $\F^\sigma_G(f_n)$ converges to $\int_G f(a) W_{1/2}(a) d\mu(a)$ in the weak operator topology of $\mc B(\Hi)$, which means that
            $$\F^\sigma_G(f) = \int_G f(a) W_{1/2}(a) d\mu(a).$$
    
        \item The set $G \times \tor$ can be equipped with the ``Heisenberg'' group law $(x,z)\cdot (y,w) = (x+y, zw\sigma(x,y))$. We denote the resulting locally compact group by $G(\sigma)$, which is a {\em central extension} of $G$. The original version of \cite[Theorem 7.1]{KleppnerLipsman72} (which applies to more general classes of groups) assumes that $G(\sigma)$ has a {\em type I} regular representation, which is the case for any abstract quantum kinematical system $(G,\sigma)$. Indeed, the quotient space $G(\sigma)/G$ can easily be identified with the group $\tor$, and the canonical Haar measure on $\tor$ is $G(\sigma)$-invariant as well. Thus, we can apply \cite[Theorem 1]{Kallman70} to conclude that $G(\sigma)$ is type I.
    \end{enumerate}
\end{rem}

In bosonic systems, one often considers another function on phase space associated to a quantum state $\rho$. It is called the Wigner function $\W = \W_\rho$, and is defined as the (symplectic) Fourier transform of the characteristic function $\chi_\rho$. This can be done in the full generality. Using the current assumption that $G$ is self-dual via the isomorphism $\Phi_\Delta$ \eqref{eq-symplectic-iso}, we can transfer the group Fourier transform $\F_G$ from \eqref{eq-group-Fourier} to get the {\em ``symplectic" group Fourier transform} on $G$
    $$\F^{\rm sym}_G: L^1(G) \to C_0(G)$$
given by
    $$\F^{\rm sym}_G(f)(a) := \int_G f(b) \overline{\Delta(a,b)}d\mu(b),\; a\in G,\; f\in L^1(G).$$
It follows from symplectic self-duality that there is (another) Haar measure $\mu^{\rm sym}_G$ on $G$ such that the map $\F^{\rm sym}_G$ extends to a unitary
    \begin{equation}\label{eq-dual-Haar-symplectic}
        \F^{\rm sym}_G: L^2(G, \mu) \to L^2(G, \mu^{\rm sym}_G).
    \end{equation}
We may call $\mu^{\rm sym}_G$ the symplectic dual Haar measure of $\mu$. We also have the corresponding Fourier inversion theorem as in \eqref{eq-Fourier-inversion}.

	\begin{defn} \label{defn-Wignerftn}
		The {\em Wigner function} $\W_\rho: G \to \Comp$ of $\rho \in \mc S^1(\Hi)$ is defined by the symplectic Fourier transform of its characteristic function $\chi_\rho \in L^2(G)$, i.e. $\W_\rho := \F^{\rm sym}_G(\chi_\rho)$.
	\end{defn}
The Wigner function $\W_\rho$ encodes the state $\rho$ in a dual manner to $\chi_\rho$. One such aspect is the following.
    \begin{prop}
        For a quantum state $\rho \in \D(\Hi)$ the Wigner function $\W_\rho$ is always real-valued and if it is integrable, then we have
            \begin{equation}\label{eq-Wigner-integral}
                \int_G \W_\rho(a)\, d\mu^{\rm sym}_G(a) = 1.
            \end{equation}
    \end{prop}
\begin{proof}
The first conclusion follows from the fact that $W_{1/2}$ is involutive (as it is normalized): ${\rm Tr}(\rho W_{1/2}(-a)^*) = {\rm Tr}(\rho W_{1/2}(a)) = \overline{{\rm Tr}(\rho W_{1/2}(a)^*)}$, $a\in G$. When $\W_\rho\in L^2(G, \mu^{\rm sym}_G)$ is also integrable, then the associated function $\chi_\rho$ must be continuous by \cite[Theorem 4.4.13]{ReiSte}, so we can safely take evaluation $\chi_\rho$. In particular, we see that $\chi_\rho(0)=1$, so by Fourier inversion we have $\int_G \W_\rho(a) d\mu^{\rm sym}_G(a) = 1$.
\end{proof}

The above Proposition (which is well known for bosonic systems) is the reason why Wigner functions are called ``pseudo-probability distributions". It is of interest to investigate the class of states whose Wigner functions are actual probability measures, equivalently, non-negative. We will focus on this theme in Section \ref{sec-Hudson}, but for now we record one useful property of characteristic/Wigner functions which follows directly from the CCR \eqref{eq-Weyl-CCR}: for $\rho\in \D(\Hi)$ we have
\begin{equation} \label{eq-chftnclifford}
    \chi_{W_{1/2}(z)^*\rho W_{1/2}(z)}(w)=\overline{\Delta(z,w)}\chi_{\rho}(w),\;\; w,z \in G
\end{equation}
\begin{equation} \label{eq-wignerclifford}
    \mathcal{W}_{W_{1/2}(z)^*\rho W_{1/2}(z)}(w)=\mathcal{W}_{\rho}(w+z),\;\; w, z\in G.
\end{equation}

\begin{rem}${}$
    \begin{enumerate}
        \item The above Wigner function exhibits similar properties to bosonic Wigner functions, but we will postpone collecting such properties until the follow-up paper \cite{GQIT2}.
        
        \item Our Wigner functions coincide with the ones from \cite{Mukunda1979}, \cite{rigas2010non} and \cite{Gro}.
    \end{enumerate}
\end{rem}

\section{Gaussian states in general quantum kinematical systems}\label{sec:Gaussian}

We first recall some necessary background on Gaussian distributions over second countable LCA groups $G$, and refer the reader to \cite{Feldman} for details.

\begin{defn} A distribution $\nu$ on $G$ is called
\begin{itemize}
    \item \textit{Gaussian} if its Fourier transform $\hat{\nu}$ on $\widehat{G}$ is of the form
    \begin{equation}\label{eq:x}\hat{\nu}(\gamma)=\la\gamma,x\ra \exp(-\varphi(\gamma)), \ \ \ \gamma\in \widehat{G},\end{equation}
    for some $x\in G$ and some non-negative continuous $\varphi: \widehat{G} \to \Real$ satisfying
    \begin{equation}\label{e:parallel}\varphi(\gamma+\gamma')+\varphi(\gamma -\gamma')=2(\varphi(\gamma)+\varphi(\gamma')), \ \ \ \gamma,\gamma'\in \widehat{G}.\end{equation}
    \item \textit{Gaussian in the sense of Bernstein}, or simply \textit{B-Gaussian} if
	\begin{equation}\label{e:BGauss}\hat{\nu}(\gamma +\gamma')\hat{\nu}(\gamma - \gamma') = \hat{\nu}(\gamma)^2|\hat{\nu}(\gamma')|^2, \ \ \ \gamma,\gamma' \in \widehat{G}.\end{equation}
\end{itemize}
\end{defn}

\begin{rem}\label{rem-B-Gaussianity}${}$
\begin{enumerate}
    \item Gaussian distributions on LCA groups were first studied by  Parthasarathy, Rao, and Varadhan \cite{Partha1963} as a generalization of Gaussian distributions on $\Real^n$. This concept has been further generalized to B-Gaussian distributions by Rukhin \cite{Rukhin1969} and by Heyer and Rall \cite{HeyerRall1972} through analogues of the Kac-Bernstein theorem on LCA groups.
    Note that if the group $G$ contains a closed subgroup homeomorphic to $\tor^2$, then we can always find a B-Gaussian distribution on $G$ which is not Gaussian \cite[Lemma 9.6]{Feldman}.
        
    \item Any non-negative continuous function $\vphi:\widehat{G}\rightarrow\bR$ satisfying (\ref{e:parallel}) is of the form
        $$\varphi(\gamma) = \psi(\gamma,\gamma),\; \gamma\in \widehat{G},$$
    where $\psi:\widehat{G}\times \widehat{G}\rightarrow\bR$ is a continuous function satisfying
    \begin{itemize}
                \item $\psi(\gamma_1, \gamma_2) = \psi(\gamma_2, \gamma_1)$,

        \item $\psi(\gamma_1 + \gamma_2, \gamma_3) = \psi(\gamma_1, \gamma_3) + \psi(\gamma_2, \gamma_3)$,
                \item $\psi(\gamma_1,\gamma_1)\ge 0$
    \end{itemize}
    for any $\gamma_1, \gamma_2, \gamma_3 \in \widehat{G}$. In particular, $\psi \in \mathrm{Hom}(\widehat{G},\mathrm{Hom}(\widehat{G},\Real))$.
    
    \item From the definition we can easily see that the Fourier transform of a Gaussian distribution $\nu$ on $G$ is fully supported, i.e. ${\rm supp}\, \nu = \wh{G}$·

\end{enumerate}

\end{rem}

We collect some properties of B-Gaussian distributions which will be useful throughout this paper. An LCA group $K$ is called a \textit{Corwin group} if $2K := \{2k: k\in K\} = K$, i.e., the doubling map is surjective.

\begin{prop}\label{prop-B-Gaussian}
Let $\nu$ be a B-Gaussian distribution on $G$ and
    $$H = {\rm supp}\,\hat{\nu} = \set{\gamma\in \widehat{G}: \hat{\nu}(\gamma)\neq 0}.$$
\begin{enumerate}
    \item The set $H$ is an open subgroup of $\widehat{G}$, whose annihilator $H^{\perp}$ is a compact Corwin subgroup of $G$.
    
    %\item ({\bf DO we need this?}) The Fourier transform $\hat{\nu}$ of $\nu$ is of the form $$\hat{\nu}(\gamma)=1_H(\gamma)l(\gamma)\exp(-\varphi(\gamma)),\;\; \gamma \in \wh{G}$$ for some continuous functions $\varphi:H\to [0,\infty)$ satisfying \eqref{e:parallel} and $l:H\to \tor$ satisfying \begin{equation}\label{eq-l}l(\gamma+\gamma')l(\gamma-\gamma')=l(\gamma)^2,\;\; \gamma,\gamma'\in \widehat{G}.\end{equation}
    
    \item
    Suppose that $G$ has no subgroup isomorphic to $\tor^2$ and $H=\widehat{G}$. Then $\nu$ is a Gaussian distribution on $G$.    
        
    \item If $G_e$, the connected component of the identity of $G$, contains at most one element of order 2, then $\nu = \nu_0 * (1_K/\mu(K))$ for a compact Corwin subgroup $K$ of $G$ and a Gaussian distribution $\nu_0$ on $G$. The mentioned hypothesis on $G$ is satisfied when $G$ is discrete or 2-regular.
    
\end{enumerate}
\end{prop}
\begin{proof}
(1) Openess of $H$ is clear, and $H$ being a subgroup is direct from \eqref{e:BGauss}. Moreover, the quotient group $\wh{G}/H$ is discrete, so that $H^\perp \cong \widehat{\wh{G}/H}$ is compact. For the Corwin property of $H^\perp$, it suffices to check that $2\gamma \in H$ implies that $\gamma\in H$ by \cite[Lemma 7.2]{Feldman}. But this is also direct from \eqref{e:BGauss}.

\vspace{0.5cm}

(2)$\&$(3) These are \cite[Lemma 9.7, Theorem 9.9]{Feldman}.
\end{proof}

We are now ready to define Gaussian states over general kinematic systems. 

\begin{defn}\label{d:Gauss} A state $\rho\in \mc D(\Hi)$ is \textit{Gaussian} (resp. \textit{B-Gaussian}) if there is a Gaussian (resp. B-Gaussian) distribution $\nu$ on $\widehat{G}$ such that $\chi_\rho = \F_{\wh{G}}(\nu)$.
\end{defn}

\begin{rem}\label{rem-normalized-is-needed}${}$
\begin{enumerate}
    \item Note that Definition \ref{d:Gauss} requires $\chi_{\rho}$ to be the Fourier transform of a (B-)Gaussian distribution on $\wh{G}$ instead of a (B-)Gaussian distribution on $G$. This difference does not show up in the bosonic system since the class of Gaussian distributions on $\Real^n$ are preserved by Fourier transform. %This (natural) choice allows us to recover several known interesting examples of states as you can see in Example
    %\ref{ex:stab}/\ref{ex:UP}.
    
    \item
    Conjugation with respect to a Weyl operator preserves (B-)gaussian states. More precisely, \eqref{eq-chftnclifford} tells us that for any $a\in G$ the state $W_{1/2}(a)^*\rho W_{1/2}(a)$ is Gaussian (resp. B-Gaussian) whenever $\rho$ is.
    
    \item Every Gaussian state is clearly a B-Gaussian state. However, the class of all B-Gaussian states is strictly larger than that of all Gaussian states in general. See Example \ref{ex:stab}/\ref{ex:UP}, Theorem \ref{t:finGauss}, Corollary \ref{t:fin-nongauss} and Proposition \ref{prop-finGauss-pure1}  below for such cases.
    
    \item In order to secure the existence of B-Gaussian states we need to focus on normalized 2-cocycles. Indeed, suppose $\rho\in \mc D(\Hi)$ is a B-Gaussian state with respect to the $\sigma$-representation $W$ where $\sigma$ is a general 2-cocycle $\sigma$ on $G$. The positivity of $\rho$ says that
        $$\chi_\rho(a)=\chi_{\rho^*}(a)=\overline{\sigma(a,-a)}\tr(\rho^* W(-a))=\overline{\sigma(a,-a)}\; \overline{\chi_{\rho}(-a)}.$$
    Being a Fourier transform of a distribution, we have $\chi_{\rho}(a)=\overline{\chi_{\rho}(-a)}, a\in G$. Therefore, we have $\sigma(a,-a)=1$ whenever $\chi_{\rho}(a)\neq 0$, i.e. on the support of $\chi_\rho$, which is an open subgroup of $G$ and non trivial in many cases. This is one reason why we require our quantum kinematical system to be equipped with normalized 2-cocycles.

\end{enumerate}

\end{rem}

For bosonic systems, Gaussianity and B-Gaussianity coincide with the usual notion of bosonic Gaussian states (see, e.g., \cite[\S 12.3.2]{Hol}) by the multivariate Kac-Bernstein theorem. See \cite{Feldman} for more details, generalizations, and further references. We now present some examples of B-Gaussian states which were already well-known in the literature under different names. To see this, first recall that for a closed subgroup $H\leq G$, its \textit{symplectic complement} is defined by
   $$H^{\Delta}:=\{z\in G\mid\Delta(z,h)=1 \ \forall \ h\in H \}.$$
We say that $H$ is \textit{isotropic} (respectively, \textit{maximally isotropic or Lagrangian}) if $H\subseteq H^{\Delta}$ (respectively, $H=H^\Delta$).

\begin{ex}\label{ex:stab} (Discrete stabilizer states) Let $d\ge 3$ be an odd integer, and consider the Weyl system $(\bZ_d^n\times\bZ_d^n, \tilde{\sigma}_{\rm can})$. For a maximally isotropic subgroup $H$ of $\bZ_d^n\times\bZ_d^n$ and $v\in\bZ_d^n\times\bZ_d^n$ the associated {\em stabilizer state} (see, e.g., \cite{Gro,HDD}) $|H,v\ra\la H,v|$ is the rank-1 projection
\begin{equation} \label{eq-stabilizerstate}
    |H,v\ra\la H,v|=\frac{1}{|H|}\sum_{h\in H}\Delta(v,h) W_{1/2}(h)=\frac{1}{d^n}\sum_{h\in H}\Delta(v,h) W_{1/2}(h).
\end{equation}
Indeed, the projection $|H,v\ra\la H,v|$ is the unique state stabilized by the abelian group $\set{\Delta(v,h)W_{1/2}(h):h\in H}$ \cite[Lemma 8]{Gro}, that is,  
    $$\Delta(v,h)W_{1/2}(h)|H,v\ra=|H,v\ra,\;\; h\in H.$$

As shown in the proof of \cite[Lemma 9]{Gro}, its characteristic function $\chi_{H,v} := \chi_{|H,v\ra\la H,v|}$ is of the form
    $$\chi_{H,v}(z)=\Delta(v,z)1_{H}(z), \; z\in \bZ_d^n\times\bZ_d^n,$$
where $1_{H}$ is the indicator function of $H$ (albeit with a different normalization from \cite{Gro}). Self-duality of finite abelian groups tells us that there is $v_0\in\bZ_d^n\times\bZ_d^n$ such that $\Delta(v,\cdot)=\la v_0,\cdot\ra$. Hence, $\chi_{H,v}$ is the Fourier transform of $\nu=\delta_{v_0}\ast (\frac{1}{|H^\perp|}1_{H^\perp})$. Now we check the condition \eqref{e:BGauss}. The character $\Delta(v,\cdot)$ clearly satisfies \eqref{e:BGauss}. Finiteness and 2-regularity of the group $\bZ_d^n\times\bZ_d^n$ imply that $H$ is also 2-regular, and consequently $1_H$ satisfies \eqref{e:BGauss}. This means that the stabilizer state $|H,z\ra\la H,z|$ is a pure B-Gaussian state. Note that $H$ is a non-trivial proper subgroup of $\bZ_d^n\times\bZ_d^n$  since $|H| = d^n$ and the Fourier transform of Gaussian distributions always have full support. Thus, we know that $|H,z\ra\la H,z|$ is not a Gaussian state.

Later we will show that pure B-Gaussian states in the $n$-qudit system are precisely the stabilizer states $|H,v\ra\la H,v|$ for some $v\in \z_d^{2n}$ and some maximally isotropic subgroup $H\leq \z_d^{2n}$, and that there are no Gaussian states (even mixed ones) over the Weyl system $(\bZ_d^n\times\bZ_d^n, \tilde{\sigma}_{\rm can})$. (See Theorem \ref{t:finGauss}, Corollary \ref{t:fin-nongauss}, Proposition \ref{prop-finGauss-pure1}, and Example \ref{ex-discrete}.) Hence, the Gaussian character of qudit stabilizer states (which belongs to the folklore) is made explicit through the Bernstein
identity (4.2) of their characteristic functions.
\end{ex}

\begin{ex}\label{ex:UP} (Minimum uncertainty states) Consider the Weyl system $(G=F\times\widehat{F}, \tilde{\sigma}_{\rm can})$ over a 2-regular LCA group $G$ such that $F$ contains a compact open 2-regular subgroup $K$. Fix $z_0\in G$. Then
$$\psi={\mu_F(K)^{-1/2}}W_{1/2}(z_0)1_{K}\in L^2(F)$$
is a minimum uncertainty state in the sense that it saturates the entropic uncertainty relation from \cite[Theorem 1.5]{OP}. The characteristic function $\chi_{\psi}:=\chi_{|\psi\ra\la\psi|}$ of $| \psi \ra \la \psi |$ then satisfies (by \eqref{eq-chftnclifford})
\begin{align*}
\chi_\psi(z)&=\mu^{-1}_F(K)\Delta(z_0,z)\chi_{|1_K\ra\la 1_K|}(z)=\mu^{-1}_F(K)\Delta(z_0,z)\overline{\la W_{1/2}(z)1_K, 1_K\ra},\;\; z\in G.\\
%&=\mu^{-1}_F(K)\tr(W_{1/2}(z_0)^*W_{1/2}(z)^*W_{1/2}(z_0)|1_K\ra\la 1_K|)\\
%&=\mu^{-1}_F(K)\tr(W_{1/2}(-z_0)W_{1/2}(-z)W_{1/2}(z_0)|1_K\ra\la 1_K|)\\
%&=\mu^{-1}_F(K)\Delta(z_0,z)\tr(W_{1/2}(-z)W_{1/2}(-z_0)W_{1/2}(z_0)|1_K\ra\la 1_K|)\\
%&=\mu^{-1}_F(K)\Delta(z_0,z)\tr(W_{1/2}(-z)W_{1/2}(z_0)^*W_{1/2}(z_0)|1_K\ra\la 1_K|)\\
%&=\mu^{-1}_F(K)\Delta(z_0,z)\la W_{1/2}(-z)1_K, 1_K\ra\\
%&=\mu^{-1}_F(K)\Delta(z_0,z)\overline{\la W_{1/2}(z)1_K, 1_K\ra},\; z\in G.
\end{align*}
Note that for $y\in K$, $y-x\in K$ if and only if $x\in K$. Hence, for $z = (x,\gamma) \in G$
\begin{align*}\la W_{1/2}(z)1_K, 1_K\ra&=\int_G\la x, \gamma\ra^{-1/2}\la y, \gamma\ra 1_K(y-x) \overline{1_{K}(y)}d\mu(y)\\
&=1_K(x)\int_K\la x, \gamma\ra^{-1/2}\la y, \gamma\ra d\mu(y)\\
%&=1_K(x)\la x, \gamma\ra^{1/2}\int_K\la\gamma,y\ra d\mu(y)\\
&=\mu_F(K)1_{K}(x)\la x, \gamma\ra^{-1/2}1_{K^{\perp}}(\gamma)
\end{align*}
By 2-regularity of $K$, $2^{-1}K=K$. Thus, for $x\in K$ and $\gamma\in K^{\perp}$
$$\la x, \gamma\ra^{-1/2}=\overline{\la 2^{-1}x, 2^{-1}\gamma\ra^2}=\overline{\la 2^{-1}x, \gamma\ra}=1,$$
and consequently
    $$\chi_\psi(z)=\Delta(z_0,z)1_{K\times K^{\perp}}(z),\; z\in G.$$
Similar to the previous example we see that $| \psi \ra \la \psi |$ is a pure B-gaussian state. 
%If, in addition, $F$ is non-compact such as the $p$-adic group $\mathbf{Q}_p$ with prime $p$, then 
On the other hand, since $K\times K^{\perp}$ is never equal to the whole phase space $G=F\times \widehat{F}$, $| \psi \ra \la \psi |$ is not a Gaussian state as before. The full description of pure B-Gaussian states in this setting will be given in Theorem \ref{thm-finGauss-pure2}.
\end{ex}

\section{Weyl systems over 2-regular groups}\label{sec-2-regular}

In this section we focus on the Weyl system $(G=F\times\widehat{F}, \tilde{\sigma}_{\rm can})$ over a 2-regular LCA group and provide a complete characterization of B-Gaussian states. By the structure theorem of van Kampen, we know that $F\cong \bR^n\times F_c$ for some LCA group $F_c$ admitting a compact open subgroup \cite[24.30]{HewittRoss}.
So we have $G\cong \bR^{2n}\times (F_c\times\wh{F_c})$. Let us write $G_c = F_c\times\wh{F_c}$ for later use. 

Our strategy is to first characteize B-Gaussian states on the Weyl system $(G_c, \tilde{\sigma}_{\rm can})$ and then use our result to show that, amongst B-Gaussian states over $G=\bR^{2n}\times G_c$, there can be no bipartite entanglement between the subsystems $\bR^{2n}$ and $G_c$. The full characterization then follows naturally.

\subsection{Systems admitting compact open subgroups} \label{sec-compactopen}
 
The main goal of this section is to establish the following theorem.

\begin{thm} \label{t:finGauss}
For a state $\rho\in\mc D (L^2(F_c))$, the following are equivalent:
\begin{enumerate}
\item $\rho$ is a B-Gaussian state on the Weyl system $(G_c, \tilde{\sigma}_{\rm can})$;
\item there exist a compact open 2-regular isotropic subgroup $H\leq G_c$ and a character $\gamma\in\widehat{H}$ such that
    \begin{equation} \label{eq-finGauss}
    \rho = \rho_{H,\gamma}:= \int_H\gamma(z)W_{1/2}(z)d\mu_{G_c}(z).
\end{equation}
Moreover, $H$ and $\gamma$ are uniquely determined.
\end{enumerate}
\end{thm}

Let us first focus on the easier direction $(2)\Rightarrow(1)$. The main step in the proof is Proposition \ref{prop-fin-square}, which requires a few preparatory lemmas. For notational simplicity we write $\sigma = \tilde{\sigma}_{\rm can}=\Delta^{1/2}$ from \eqref{eq-normalization}.

    \begin{lem} \label{lem-fin-symplectic-complement}
    Let $H$ be a compact open 2-regular subgroup of $G_c$.
    \begin{enumerate}
    
    \item For $z\in G$ we have $z\in H^{\Delta}$ if and only if $\sigma(z,h)=1$ for all $h\in H$.
    
    \item For $z\in G$, $\displaystyle \int_H\sigma(z,z')d\mu_{G_c}(z')= \begin{cases}\mu(H), & z\in H^{\Delta}\\ 0, & \text{otherwise} \end{cases}$.
    \end{enumerate}
    \end{lem}

\begin{proof} (1) If $z\in H^{\Delta}$, then for all $h\in H$
$$\sigma(z,h)=\Delta^{1/2}(z,h)=\Delta(2^{-1}z,2^{-1}h)^2=\Delta(z,2^{-1}h)=1$$
as $2^{-1}H\subseteq H$. Conversely, if $\sigma(z,h)=1$ for all $h\in H$ then again by definition of the normalization
$$\Delta(z,h)=\Delta(2^{-1}z,2^{-1}(2h))^2=\sigma(z,2h)=1, \ \ \ h\in H.$$
(2) This is from (1) and \eqref{eq-char-integral}.
\end{proof}

\begin{lem}\label{lem-compactopen} Let $H$ be a compact open subgroup of $G_c$. Then $H^\Delta$ is a compact open subgroup of $G_c$. Morever, if $H$ is 2-regular then so is $H^\Delta$.
\end{lem}

\begin{proof} Since $H$ is open, the quotient $G_c/H$ is discrete, so that $H^\perp \cong \widehat{G_c/H}$ is compact. Since $H$ is compact, the dual $\wh{H}$ is discrete, so that $H^\perp$ is open from $\widehat{G_c}/H^{\perp}\cong\widehat{H}$. Thus, $H^\perp$ is a compact open subgroup of $\wh{G_c}$. But $H^\perp=\Phi_\Delta(H^\Delta)$ via the isomorphism $\Phi_{\Delta}:G_c\rightarrow\widehat{G_c}$, implying that $H^\Delta$ is compact open in $G_c$.

For the final statement, let $z\in H^\Delta$. Then for all $h\in H$, 
$$\Delta(2^{-1}z,h)=\Delta(2^{-1}z,2^{-1}h)^2=\sigma(z,h),$$
so the result follows from Lemma \ref{lem-fin-symplectic-complement} (1).
\end{proof}

\begin{prop} \label{prop-fin-square}
The element $\rho_{H,\gamma}$ from \eqref{eq-finGauss} is a self-adjoint operator satisfying the relation $\rho_{H,\gamma}^2=\mu_{G_c}(H)\rho_{H\cap H^{\Delta},\gamma}$.
\end{prop}
\begin{proof}
Self-adjointness comes from $W_{1/2}(z)^*=W_{1/2}(-z)$, $\overline{\gamma(z)}=\gamma(-z)$. Since the above integral (\ref{eq-finGauss}) is WOT-convergent, and multiplication is separately WOT-continuous, we have
\begin{align*}
    \rho_{H,\gamma}^2 &=\int_H\int_H\gamma(z)\gamma(z')W_{1/2}(z)W_{1/2}(z')d\mu_{G_c}(z)d\mu_{G_c}(z')\\
    &=\int_H\int_H\gamma(z+z')\sigma(z,z')W_{1/2}(z+z')d\mu_{G_c}(z)d\mu_{G_c}(z')\\
    &=\int_H\int_H\gamma(z')\sigma(z,z'-z)W_{1/2}(z')d\mu_{G_c}(z)d\mu_{G_c}(z')\\
    &=\int_H\int_H\gamma(z')\sigma(z,z')W_{1/2}(z')d\mu_{G_c}(z)d\mu_{G_c}(z')\\
    &=\int_H\gamma(z')\bigg(\int_H\sigma(z,z')d\mu_{G_c}(z)\bigg)W_{1/2}(z')d\mu_{G_c}(z')\\
    &=\mu_{G_c}(H)\int_{H\cap H^{\Delta}}\gamma(z')W_{1/2}(z')d\mu_{G_c}(z')\\
    &=\mu_{G_c}(H)\rho_{H\cap H^{\Delta},\gamma},
\end{align*}
where the second last equality is Lemma \ref{lem-fin-symplectic-complement} (2). Note that $H\cap H^\Delta$ is again a compact open 2-regular subgroup by Lemma \ref{lem-compactopen}, so the notation $\rho_{H\cap H^{\Delta},\gamma}$ is justified.
\end{proof}

\begin{proof}[Proof of Theorem \ref{t:finGauss}] $(2)\Rightarrow(1)$: Since $H$ is isotropic, $H=H\cap H^{\Delta}$, so Proposition \ref{prop-fin-square} implies $\rho_{H,\gamma}^2=\mu_{G_c}(H)\rho_{H,\gamma}$. By the spectral theorem for compact operators, 
$\rho_{H,\gamma}$ is positive with $\mathrm{spec}(\rho_{H,\gamma})=\{0,\mu_{G_c}(H)\}$. Moreover, as $\rho_{H,\gamma}=\mc{F}^\sigma_{G_c}(\gamma1_H)$, by Lemma \ref{prop-Fourier-char} and injectivity of $\mc{F}^\sigma_{G_c}$, we have $\chi_\rho=\gamma1_H$. In particular, $\mathrm{Tr}(\rho_{H,\gamma})=\chi_\rho(0)=1$, so $\rho_{H,\gamma}$ is a state. Its characteristic function satisfies the Bernstein identity \eqref{e:BGauss}: 
\begin{align*}\chi_\rho(z+z')\chi_\rho(z-z')&=\gamma(z+z')\gamma(z-z')1_H(z+z')1_H(z-z')\\
&=\gamma(z)^2|\gamma(z')|^21_H(z)1_H(z'),
\end{align*}
where the last equality uses 2-regularity of $H$ to show $z+z',z-z'\in H$ if and only if $z,z'\in H$. On the other hand, $\chi_{\rho}=\gamma 1_H$ is continuous and positive definite since $\gamma\in \widehat{H}$ and  $H$ is a compact open subgroup \cite{We40,Po40,Ra40}. Consequently, $\chi_{\rho}$ is the Fourier transform of a B-Gaussian distribution, and thus $\rho_{H,\gamma}$ is a B-Gaussian state.
\end{proof}

The other, more involved direction of the proof of  Theorem \ref{t:finGauss}, requires additional preparations. The major step, Proposition \ref{p:singular}, concerns the singularity of Gaussian distributions in our setting, and is based on \cite[Proposition 3.14]{Feldman}. We begin with a few general lemmas. Recall that, in this paper, all LCA groups are assumed to be second countable (hence metrizable).

\begin{lem}\label{lem-Ge} Let $G$ be an LCA group, and let $H$ be an open subgroup of $G$. Then $G_e\leq H$, where $G_e$ is the connected component of the identity in $G$.
\end{lem}

\begin{proof} If $\pi:G\twoheadrightarrow G/H$ denotes the canonical quotient map, then $\pi(G_e)$ is a connected subgroup of the discrete group $G/H$. Hence, the group $\pi(G_e)$ is trivial, which implies $G_e\leq H$.
\end{proof}

\begin{lem}\label{l:pathconn} Let $G$ be a 2-regular LCA group admitting a compact open subgroup. Then, any path connected closed subgroup of $G$ must be trivial.
\end{lem}

\begin{proof} Let $H$ be a path connected closed subgroup of $G$. Together with second countability, we know that $H \cong \Real^n\times\bT^m$ for some $n\geq 0$ and $m\leq\aleph_0$ \cite[8.27]{Armacost}.

First, connectedness of $H$ implies $H\leq G_e$. Since $G$ admits a compact open subgroup, Lemma \ref{lem-Ge} implies that $G_e$ compact. Hence, $H$ is compact, which forces $n=0$.

Second, since $G$ is 2-regular, the doubling map $x\mapsto 2x$ is injective on $H$. Since this is false for any non-trivial power of $\bT$, we must also have $m=0$. Thus, $H$ is trivial.
\end{proof}

\begin{prop}\label{p:singular} Let $G$ be a non-discrete 2-regular LCA group with a compact open subgroup. Then any Gaussian distribution $\nu$ on $G$ is singular with respect to the Haar measure $\mu$ on $G$. 
\end{prop}

\begin{proof} We may assume that $\nu$ is \textit{symmetric} (meaning $x=0$ in \eqref{eq:x}) by translation. Then the support $C$ of $\nu$ is a connected closed subgroup of $G$ \cite[Proposition 3.6]{Feldman}. By Lemma \ref{l:pathconn}, if $C$ were path connected, it would be trivial, in which case $\nu=\delta_{e}$ is singular (as $G$ is not discrete).

Suppose $C$ is not path connected. We know by \cite[Proposition 3.8, Proposition 3.11]{Feldman} that there is a path connected Polish group $L$ (not necessarily locally compact), a continuous homomorphism $p:L\rightarrow C$, and a distribution $\nu_L$ on $L$ such that $\nu=p(\nu_L)$ (push-forward measure). Hence, $\nu$ is concentrated on the subgroup $p(L)$. We claim that $p(L)$ is Borel with $\mu(p(L)) = 0$, which gives the singularity of $\nu$.

First note that $p(L)$ is the image of the induced map $\tilde{p}:L/\mathrm{Ker}(p) \to C$, which is injective. Since $L/\mathrm{Ker}(p)$ is a Polish group and $C$ is metrizable, it follows that $p(L)$ is Borel by \cite[Corollary A.7]{Takesaki}.

Now suppose, by way of contradiction, that $\mu(p(L))>0$. Since $G$ admits a compact open subgroup, $G_e$ is compact by Lemma \ref{lem-Ge}. Hence, $C\leq G_e$ is compact, forcing $0<\mu(p(L))\leq\mu(C)<\infty$, which means that $p(L) = p(L)-p(L)$ contains a neighborhood of the identity of $G$ by \cite[20.17]{HewittRoss}. Thus, $p(L)$ is a subgroup of $G$ with non-empty interior, hence clopen. From the definition of the support we then have $p(L) = C$, which implies that $C$ is path connected; contradiction.
\end{proof}

We are now ready to finish the proof of the main result of this subsection. Recall that $G_c = F_c\times\wh{F_c}$ where $F_c$ is an LCA group admitting a compact open subgroup.

\begin{proof}[Proof of Theorem \ref{t:finGauss}] 

$(1)\Rightarrow(2)$: Suppose $\rho\in \D(L^2(F_c))$ is a B-Gaussian state. Then $\chi_\rho$ is the Fourier transform of a B-Gaussian distribution on $\wh{G}_c$. Thanks to 2-regularity and Proposition \ref{prop-B-Gaussian} it is of the form $\chi_\rho = \mc{F}_{\wh{G}_c}(\nu)|_{K^\perp}1_{K^\perp}$
for a compact Corwin subgroup $K\leq\wh{G}_c$ and a Gaussian distribution $\nu$ on $\wh{G}_c$.
Since $K^\perp$ is open and $\chi_\rho\in L^2(G_c)$ we know that $\mc{F}_{\wh{G}_c}(\nu)|_{K^\perp} \in L^2(K^\perp)$. We claim that
    $$(*) \;\; \mc{F}_{\wh{G}_c}(\nu)|_{K^\perp} = \mc{F}_{\wh{G}_c/K}(\nu_K)$$
for some Gaussian distribution $\nu_K$ on $\wh{G}_c/K$. Supposing $(*)$ holds, the measure $\nu_K\in M(\wh{G}_c/K)$ has square-integrable Fourier transform and so must be absolutely continuous with respect to the Haar measure on $\wh{G}_c/K$ (with square-integrable Radon-Nikodym derivative, by the Plancherel theorem). This forces $K$ to be open. If not, the group $\wh{G}_c/K$ is non-discrete, and we can appeal to Proposition \ref{p:singular} to get the contradiction that $\nu_K$ is also singular with respect to the Haar measure on $\wh{G}_c/K$. Note that $\wh{G}_c/K$ satisfies the assumption of Proposition \ref{p:singular}: $K$ is a Corwin subgroup of a 2-regular group, so it is automatically 2-regular. Together with 2-regularity of $\wh{G}_c$, it follows that $\wh{G}_c/K$ is 2-regular. Also, since $\wh{G}_c$ contains a compact open subgroup, so too does $\wh{G}_c/K$ as the canonical quotient map $\pi:\wh{G}_c\rightarrow\wh{G}_c/K$ is open (see, e.g., \cite[5.26]{HewittRoss}). Thus, $K$ is open, and the Gaussian disctribution $\nu_K$ is supported on a connected subset (\cite[Proposition 3.6]{Feldman}) of the discrete group $\wh{G}_c/K$, so that $\nu_K=\delta_{\gamma_0+K}$ for some $\gamma_0\in\wh{G}_c$.
But then  
$$\chi_\rho=\mc{F}_{\wh{G}_c/K}(\nu_K)1_{K^{\perp}}=\gamma_0^{-1}1_{K^{\perp}},$$
so letting $H=K^\perp$, and $\gamma=\gamma_0^{-1}|_{H}$, we see that $H$ is a compact open 2-regular subgroup of $G$, $\gamma\in\wh{H}$, and $\chi_\rho=\gamma1_H$. Thus, $\rho=\rho_{H,\gamma}$ as in \eqref{eq-finGauss}.

Let us go back to the claim $(*)$. Viewing $C_b(\wh{G}_c/K)\subseteq C_b(\wh{G}_c)$ in the canonical fashion (as functions which are constant on the cosets of $K$), restriction to $C_b(\wh{G}_c/K)$ induces a probability preserving map from $M(\wh{G}_c)$ to $M(\wh{G}_c/K)$. Write $\nu_K$ for the image of $\nu$ under this map. Then, for all $f\in C_b(\wh{G}_c/K)$,
$$\la f,\nu_K\ra_{(C_b(\wh{G}_c/K),M(\wh{G}_c/K))}=\la f,\nu\ra_{(C_b(\wh{G}_c),M(\wh{G}_c))}.$$ Consequently, $\nu_K$ is a Gaussian distribution on $\wh{G}_c/K$ and for $z\in\widehat{(\wh{G}_c/K)}\cong K^\perp\leq G$,
\begin{align*}\mc{F}_{\wh{G}_c/K}(\nu_K)(z)&=\int_{\wh{G}_c/K}\overline{\la z,\gamma+K\ra}d\nu_K(\gamma+K) =\la z^{-1},\nu_K\ra_{(C_b(\wh{G}_c/K),M(\wh{G}_c/K))}\\
&=\la z^{-1},\nu\ra_{(C_b(\wh{G}_c),M(\wh{G}_c))}\\
&=\mc{F}_{\wh{G}_c}(\nu)|_{K^\perp}(z).
\end{align*}

It remains to show that $H$ is isotropic. Uniqueness follows from (twisted) Fourier inversion.
If we denote $H_0=H\cap H^{\Delta}$, then $\rho^2=\mu_{G_c}(H)\rho_{H_0,\gamma}$ by Proposition \ref{prop-fin-square}. Since $H_0$ is compact, open, isotropic and 2-regular, $\rho_{H_0,\gamma}$ is a B-Gaussian state with $\rho_{H_0,\gamma}^2=\mu_{G_c}(H_0)\rho_{H_0,\gamma}$.
The eigenvalues of $\rho_{H_0,\gamma}$ are therefore 0 and $\mu_{G_c}(H_0)$. Since $\tr(\rho_{H_0,\gamma})=1$, the eigenvalue $\mu_{G_c}(H_0)$ has multiplicity $\mu_{G_c}(H_0)^{-1}$, implying that $\rho$ has the eigenvalue $\mu_{G_c}(H)^{1/2}\mu_{G_c}(H_0)^{1/2}$ with the same multiplicity.
From the condition $\tr\rho=1$ we get 
$$\mu_{G_c}(H)^{1/2}\mu_{G_c}(H_0)^{1/2}\mu_{G_c}(H_0)^{-1}=1,$$
which implies $\mu_{G_c}(H)=\mu_{G_c}(H_0)$. If $z\in H\backslash H_0=H\cap H_0^c$, then there is an open neighbourhood $U$ of $z$ in $H\backslash H_0$. But then 
$$\mu_{G_c}(H)\geq\mu_{G_c}(H_0\cup U)=\mu_{G_c}(H_0)+\mu_{G_c}(U)>\mu_{G_c}(H_0),$$
contradiction. Thus, $H=H_0$ is isotropic. 
\end{proof}

\begin{rem}\label{rem-2reg}${}$
    \begin{enumerate}
    
        \item If $G_c$ admits no compact open \textit{2-regular} subgroup, then Theorem \ref{t:finGauss} tells us that there are no B-Gaussian states.
        
        \item There are 2-regular LCA groups with no non-trivial, proper closed 2-regular subgroups. For example, take the 2-adic rationals $\mathbf{Q}_2$. Then $\mathbf{Q}_2$ is 2-regular as it is a field. However, if $H$ is a non-trivial 2-regular closed subgroup of $\mathbf{Q}_2$ then necessarily $2^{-1}H=H$. But then $2^{-n}H=H$ for all $n\in\bN$. Pick $x\in H$ with $|x|_2>0$. Then $(2^{-n}x)$ is a sequence in $H$ with $|2^{-n}x|_2=2^n|x|_2\rightarrow\infty$ as $n\rightarrow\infty$. Hence, $H$ is not bounded and therefore not compact. However, every proper closed subgroup of $\mathbf{Q}_2$ is compact (and open) \cite[Corollary 9]{RS}, so $H=\mathbf{Q}_2$. A similar argument shows that any closed 2-regular subgroup of $\mathbf{Q}_2^n$ is not compact. In particular, there is no B-Gaussian state over the 2-adic Weyl system $\mathbf{Q}_2^n\times \widehat{\mathbf{Q}_2^n}\cong \mathbf{Q}_2^{2n}$.
    \end{enumerate}

\end{rem}

The following Corollary is a reason for us to consider B-Gaussian states instead of Gaussian states. 

\begin{cor}\label{t:fin-nongauss} There is no Gaussian state in the Weyl system $(G_c=F_c\times\widehat{F_c}, \tilde{\sigma}_{\rm can})$ unless $F_c$ is trivial. 
\end{cor}
\begin{proof}
If $\rho$ is a Gaussian state, then it is B-Gaussian, so $\rho=\rho_{H,\gamma}$ for $H,\gamma$ as in Theorem \ref{t:finGauss}, and $\chi_{\rho}=\gamma1_H$. Since every Gaussian state has non-vanishing characteristic function, we have $G_c=H$.
%{\color{blue}(\textbf{Original proof}) Since $G\cong \widehat{G}$ (via the symplectic self-duality $\Phi_{\Delta}$) and $G=H$ is compact, $G$ is also discrete, and hence $G$ is finite. Now Lemma \ref{lem-UP} \tcr{(does this lemma have to be introduced first?)} says that $|G|^2=|H|^2\leq |H||H^{\Delta}|=|F|^2=|G|$ for the finite group $G$. This contradicts $G=H\subset H^{\Delta}$ unless $G$ is trivial.}\\
However, isotropy of $H$ and non-degeneracy of the symplectic form $\Delta$ implies that
    $$G_c=H\subset H^{\Delta}=G^{\Delta}_c=\set{0}.$$
\end{proof}

\begin{rem}
Based on the characterization of B-Gaussian states we can easily determine their von Neumann entropy. Indeed, in the proof of Theorem \ref{t:finGauss}, we saw that the non-zero spectrum of $\rho_{H,\gamma}$ is $\mu_{G_c}(H)$ with multiplicity $\mu_{G_c}(H)^{-1}$. It follows that
    \begin{equation}\label{eq-entropyformula}
        S(\rho_{H,\gamma}) = \log(\mu_{G_c}(H)^{-1}).   
    \end{equation}
\end{rem}

\begin{ex} \label{ex-discrete}
When $F$ is a 2-regular finite abelian group (here 2-regularity equivalent to $F$ having odd cardinality), our Haar measure on $G=F\times \widehat{F}$ satisfying Theorem \ref{thm-twistedPlancherel} is $\mu(\cdot)=|\cdot|/|F|$, where $|\cdot|$ denotes the cardinality. Therefore, the B-Gaussian state $\rho_{H,\gamma}$ can be written as 
    $$\rho_{H,\gamma}=\frac{1}{|F|}\sum_{z\in H}\gamma(z)W_{1/2}(z).$$
Moreover, we will see later that $\rho_{H,\gamma}$ is pure if and only if $H$ is maximally isotropic, or equivalently, $|H|=|F|$ (Lemma \ref{lem-Lagrangian}, Proposition \ref{prop-finGauss-pure1}). In particular, if $G=\z_d^n\times \widehat{\z_d^n}\cong \z_d^n\times {\z_d^n}$ with $d$ odd, we have $\rho_{H,\gamma}=|H,v\ra\la H,v|$ (represented as in \eqref{eq-stabilizerstate}) for some $v\in G$, by  symplectic duality. Therefore, \textit{pure B-Gaussian states over $\z_d^n\times {\z_d^n}$ coincide with stabilizer states of $n$-qudit systems}. 
%That is, the Gaussian character of stabilizer states (which belongs to the folklore) is made explicit through the Bernstein identity (\ref{e:BGauss}) of their characteristic functions.

\end{ex}

\begin{rem} \label{rem-stabilizer}

From the phase space perspective, the starting point of the stabilizer formalism of quantum error correction \cite{CRSS,Got2} is an isotropic subgroup $H$ of $G=\z_2^{2n}\cong \z_2^n\times \widehat{\z_2^n}$. The same idea works for more general phase groups $G=F_c\times \widehat{F_c}$: for a compact open 2-regular isotropic subgroup $H\leq G$ and a character $\gamma\in \widehat{H}$, one can encode information in the subspace of the system Hilbert space $L^2(F_c)$ which is stabilized/fixed by the action of (the abelian group) $\mathcal{S}=\set{\gamma(h)W_{1/2}(h):h\in H}$. The B-Gaussian state $\rho_{H,\gamma}$ is precisely the normalized projection onto the \textit{stabilizer subspace} 
    $$C(\mathcal{S})=\set{\psi\in L^2(F_c): s|\psi\ra=|\psi\ra \text{ for all } s\in \mathcal{S}}.$$
Indeed, 
$$P:={\mu_G(H)}^{-1}\rho_{H,\gamma}={\mu_G(H)}^{-1}\int_H\gamma(h')W_{1/2}(h')d\mu_G(h')$$
satisfies $P^2=P\geq 0$ (Proposition \ref{prop-fin-square}), so $P$ is an orthogonal projection. Moreover, we can show that $\gamma(h)W_{1/2}(h)P=P$ for all $h\in H$ from the definition (since $h\mapsto \gamma(h)W_{1/2}(h)$ is a group homomorphism), so ${\rm Ran}(P)$ is contained in $C(\mathcal{S})$. Finally, every vector $|\psi\ra$ stabilized by $\mathcal{S}$ clearly satisfies $P|\psi\ra=|\psi\ra$, which means that $C(\mathcal{S})\subset {\rm Ran}(P)$.

\end{rem}

\begin{ex} 
%The field $\mathbf{Q}_p$ of $p$-adic numbers is a standard example of $2$-regular non-compact abelian groups containing compact open subgroups. See \cite{Folland-book} for more concrete descriptions. 
In Zelenov's (relatively recent) papers \cite{Z,Z2}, Gaussian states on $L^2(\textbf{Q}_p)$ were defined by $\chi_\rho$ being the indicator function of a \textit{lattice} $L\subseteq \mathbf{Q}_p\times\mathbf{Q}_p$ (multiplied by a suitable character on $L$). By a lattice, they mean a rank-2 free $\mathbf{Z}_p$-submodule of $\mathbf{Q}_p\times\mathbf{Q}_p$, where $\mathbf{Z}_p=\{x\in\mathbf{Q}_p\mid |x|_p\leq 1\}$ is the ring of $p$-adic integers. Concretely, this means that there exist $\textbf{Z}_p$-linearly independent $z_1,z_2\in\mathbf{Q}_p\times\mathbf{Q}_p$ such that $L=\mathbf{Z}_p z_1\oplus\mathbf{Z}_p z_2$. Their Gaussian terminology was justified through the observation that such indicator functions are eigenfunctions of the symplectic Fourier transform.

Let us check that \textit{Gaussian states in the sense of Zelenov coincide with B-Gaussian states.} %reduce to such ``lattice states'' in the $p$-adic context.
To this end, let $G=\mathbf{Q}_p^n\times\mathbf{Q}_p^n$ with $p$ an odd prime (so that $G$ possesses B-Gaussian states, Remark \ref{rem-2reg}(2)). We equip $G$ with the metric induced by the norm
$$\norm{z}=\max_{1\leq i\leq 2n}|z_i|_p, \ \ \ z=(z_1,...,z_{2n})\in G.$$
Note that the closed unit ball of $G=\mathbf{Q}_p^{2n}$ in this norm is $\mathbf{Z}_p^{2n}$.

Let $\rho$ be a B-Gaussian state on $L^2(\mathbf{Q}_p^n)$. Since $G$ is 2-regular and admits a compact open subgroup, by Theorem \ref{t:finGauss} there exist a compact open (2-regular) isotropic subgroup $H$ and a character $\gamma\in\widehat{H}$ such that $\rho=\rho_{H,\gamma}$. Note that any closed subgroup of $G$ is automatically 2-regular since it is a $\mathbf{Z}_p$-submodule and $\frac{1}{2}\in\mathbf{Z}_p$ for odd primes $p$. By compactness there exists $N\in\bN$ for which $H\subseteq p^{-N}\mathbf{Z}_p^{2n}$. Hence, $p^{N}H$ is a $\mathbf{Z}_p$-submodule of the free module $\mathbf{Z}_p^{2n}$. Since $\mathbf{Z}_p$ is a principle ideal domain, $p^{N}H$ is free of rank at most $2n$. In addition, $H$, and therefore $p^{N}H$ is open in $\mathbf{Q}_p^{2n}$, so there is some $k\in \bN$ such that $B_{\leq p^{-k}}(0)^{2n}\subseteq p^{N}H$, where $B_{\leq p^{-k}}(0)=p^k\mathbf{Z}_p$ is the clopen ball of radius $p^{-k}$ in $\mathbf{Q}_p$. It follows that $p^{k}e_i\in p^{N}H$, where $e_i$ are the standard basis vectors of $\mathbf{Q}_p^{2n}$, so $p^{N}H$ contains at least $2n$ independent elements. Therefore, the rank of $p^{N}H$ is $2n$, implying the existence of $\mathbf{Z}_p$-independent $h_1,...,h_{2n}\in H$ for which $H=\mathrm{span}_{\mathbf{Z}_p}\{p^{-N}h_1,...,p^{-N}h_{2n}\}$. Hence, $H$ is a free $\mathbf{Z}_p$-module of rank $2n$ inside $\mathbf{Q}_p^{2n}$, that is, a lattice.

%By openness of $H$ we also have $\mathrm{span}_{\mathbf{Q}_p}H=\mathbf{Q}_p^{2n}$ so that $H$ is a lattice in $\mathbf{Q}_p^{2n}$ in the usual sense ({\bf reference!}). 

Conversely, let $\rho$ be a Gaussian state on $L^2(\textbf{Q}_p^n)$ in the sense of Zelenov associated to a lattice $L$. To prove $\rho$ is B-Gaussian, it suffices to show that $L$ is a 2-regular compact open isotropic subgroup of $\textbf{Q}_p^n\times \textbf{Q}_p^n$. Indeed, $L= \bigoplus_{i=1}^{2n} \textbf{Z}_p z_i$ for some independent $z_1,\ldots, z_{2n}\in\textbf{Q}_p^{2n}$, and therefore, 
    $$(p^N\textbf{Z}_p)^{2n}=\bigoplus_{i=1}^{2n}\textbf{Z}_p(p^Ne_i) \subset L\subset \bigoplus_{i=1}^{2n}\textbf{Z}_p(p^{-N}e_i) = (p^{-N}\textbf{Z}_p)^{2n}$$
for sufficiently large $N$. Since $\bigoplus_{i=1}^{2n} \textbf{Z}_p z_i$ is clearly closed in $\textbf{Q}_p^n\times \textbf{Q}_p^n$, this inclusion explains that $L$ is compact and open. The closedness again implies that $L$ is 2-regular as before. Now we apply the same argument as in the proof of Theorem \ref{t:finGauss} for the direction $(1)\Rightarrow(2)$ to show that $\chi_{\rho}=1_{L}$ is a characteristic function of a state only if $L$ is isotropic.

\end{ex}

\subsection{General 2-regular systems}

Let us go back to the Weyl system $(F\times\widehat{F}, \tilde{\sigma}_{\rm can})$ over a general 2-regular LCA group, where $F \cong \Real^n \times F_c$ with $F_c$ admitting a compact open subgroup.

\begin{thm}\label{p:noent} Every $B$-Gaussian state in the Weyl system $(G = F\times\widehat{F}, \tilde{\sigma}_{\rm can})$ is of the form $\rho_n\ten \rho_c$, where $\rho_n$ and $\rho_c$ are $B$-Gaussian states in the Weyl system $(\bR^n\times\widehat{\bR^n}, \tilde{\sigma}_{\rm can})$ and $(G_c = F_c\times \widehat{F_c}, \tilde{\sigma}_{\rm can})$, respectively.
\end{thm}

\begin{proof} Suppose $\rho\in\mc D(L^2(F))$ is B-Gaussian. Proposition \ref{prop-B-Gaussian}(1) implies that the support of $\chi_\rho$ is an open subgroup $H$ of $F\times\wh{F}\cong\bR^{2n}\times(F_c\times\wh{F_c})$. Thus, $H\cong \bR^{2n}\times K$ for an open subgroup $K$ of $F_c\times\wh{F_c}$. A straightforward calculation shows that the reduced state
$(\tr\ten\id)\rho\in\mc{D}(L^2(F_c))$ satisfies
$$\chi_{(\tr\ten\id)\rho}=\chi_\rho|_{\{0\}\times G_c},$$
so that $(\tr\ten\id)\rho$ is a B-Gaussian state over the Weyl system $(G_c, \tilde{\sigma}_{\rm can})$ with $\chi_{(\tr\ten\id)\rho}$ supported on $K$, which must be compact thanks to Theorem \ref{t:finGauss}.

Now we apply Proposition \ref{prop-B-Gaussian}(3) to get $\chi_\rho=1_H\gamma \exp(-\vphi)$ for some $\gamma \in \wh{H}$ and some non-negative continuous $\varphi: H \to \Real$ satisfying \eqref{e:parallel}. Let $\psi:H\times H\rightarrow\bR$ be the continuous biadditive form associated to $\varphi$ in Remark \ref{rem-B-Gaussianity}(2).
We therefore obtain a continuous homomorphism
$$H\ni z\mapsto \psi(z,\cdot)\in\mathrm{Hom}(H,\bR),$$
where $\mathrm{Hom}$ denotes the set of continuous homomorphisms. Since $H\cong \bR^{2n}\times K$, the above homomorphism can be regarded as an element of 
$$\mathrm{Hom}(\bR^{2n}\times K,\mathrm{Hom}(\bR^{2n}\times K,\bR)),$$
which, by commutativity of $\bR$, identifies canonically with the product group
\begin{align*}&\mathrm{Hom}(\bR^{2n},\mathrm{Hom}(\bR^{2n},\bR))\times\mathrm{Hom}(K,\mathrm{Hom}(\bR^{2n},\bR))\\
\times&\mathrm{Hom}(\bR^{2n},\mathrm{Hom}(K,\bR))\times\mathrm{Hom}(K,\mathrm{Hom}(K,\bR)).\end{align*}
Under this identification, we may write
$$\psi((x,y),(x',y'))=\bigg\la\begin{bmatrix} A & B\\ C & D\end{bmatrix}\begin{bmatrix} x \\ y\end{bmatrix},\begin{bmatrix} x' \\ y'\end{bmatrix}\bigg\ra,\ \ \ x,x'\in\bR^{2n}, \ y,y'\in K.$$
where $A\in\mathrm{Hom}(\bR^{2n},\mathrm{Hom}(\bR^{2n},\bR))\cong M_{2n}(\bR)$, $B\in \mathrm{Hom}(K,\mathrm{Hom}(\bR^{2n},\bR))\cong\mathrm{Hom}(K,\bR^{2n})$, $C\in\mathrm{Hom}(\bR^{2n},\mathrm{Hom}(K,\bR))$ and $D\in\mathrm{Hom}(K,\mathrm{Hom}(K,\bR))$. Since $K$ is compact, $\mathrm{Hom}(K,\bR^m)=\{0\}$ for any $m\in\bN$. Thus, $B=C=D=0$, and we have
$\psi((x,y),(x',y'))= \la Ax,x'\ra,\, x,x'\in\bR^{2n}, y,y'\in K$ and consequently
    $$\varphi((x,y)) = \psi((x,y),(x,y)) = \la Ax, x \ra,\; x\in\bR^{2n}, y\in K.$$

Since $\gamma\in\wh{H}\cong \wh{\bR^{2n}\times K}=\wh{\bR^{2n}}\times\wh{K}$, we may write $\gamma=\gamma_n \times \gamma_c$ with $\gamma_n\in\wh{\bR^{2n}}$, $\gamma_c\in\wh{K}$. Putting things together, we see that
$$\chi_\rho(x,y)=1_K(y)\gamma_c(y)\gamma_n(x)\exp(-\la Ax,x\ra)=\chi_n(x)\chi_c(y), \ \ \ x\in\bR^{2n}, \ y\in K,$$
where $\chi_n=\chi_\rho|_{\bR^{2n}\times\{0\}}=\chi_{(\id\ten\tr)\rho}$ and $\chi_c=\chi_\rho|_{\{0\}\times K}=\chi_{(\tr\ten\id)\rho}$ are the characteristic functions of B-Gaussian sates in $\rho_n\in\D(L^2(\bR^{n}))$ and $\rho_c\in\D(L^2(F_c))$, respectively. By uniqueness of characteristic functions, it follows that $\rho=\rho_n\ten\rho_c$, where $\rho_n=(\id\ten\tr)\rho$ and $\rho_c=(\tr\ten\id)\rho$. 
\end{proof}

\begin{rem}
Theorem \ref{p:noent} shows that there is a topological obstruction for $B$-Gaussian states over the Weyl system $(F\times\widehat{F}, \tilde{\sigma}_{\rm can})$ with $F \cong \Real^n \times F_c$ to have bipartite entanglement with respect to the decomposition $L^2(F)\cong L^2(\bR^n)\ten L^2(F_c)$. A similar separability phenomenon is known to hold for minimizers of the entropic uncertainty principle over LCA groups \cite{OP}.
\end{rem}

%\tcr{Add corollaries: convexity, symmetric bicharacter result, etc...}

\subsection{Pure Gaussian states} \label{sec-pureGaussian}

Based on our characterization (Theorem \ref{p:noent}), every $B$-Gaussian state in a 2-regular Weyl system $(F\times\widehat{F}, \tilde{\sigma}_{\rm can})$ is of the form $\rho_n\ten \rho_c$, where $\rho_n$ and $\rho_c$ are $B$-Gaussian states in the Weyl systems $(\bR^n\times\widehat{\bR^n}, \tilde{\sigma}_{\rm can})$ and $(G_c, \tilde{\sigma}_{\rm can})$, respectively.
Since a product state is pure if and only if each component is pure, and the purity of bosonic Gaussian states has been characterized (see \cite[Section 3]{AGI07}, for example), the characterization of pure B-Gaussian states reduces to the case of $\rho_c$. By Theorem \ref{t:finGauss}, it is of the form $\rho_{H,\gamma}$ for some compact open 2-regular isotropic subgroup $H\leq G_c$ and a character $\gamma\in\widehat{H}$. In Proposition \ref{prop-finGauss-pure1} we will prove that $\rho_{H,\gamma}$ is pure if and only if $H$ is maximally isotropic. Moreover, we show that every pure B-Gaussian state is determined (up to a Weyl translation) by a symmetric bicharacter (Theorem \ref{thm-finGauss-pure2}). We begin with some preliminary results.

\begin{lem}\label{lem-UP} For any compact open subgroup $H\leq G_c$ we have
\begin{equation}\label{eq:UP}\mu(H)\mu(H^\Delta)=1,\end{equation}
where $\mu = \mu_{G_c}$.
\end{lem}

\begin{proof} Let $\Phi_\Delta$ be the (canonical) symplectic self-duality on $G_c$. By uniqueness of Haar measures, there exists $c>0$ for which $\Phi_\Delta(\mu)=c\mu_{\wh{G}_c}$, where $\Phi_\Delta(\mu)$ is the push-forward measure.
Since $H^\perp=\Phi_\Delta(H^\Delta)$, by the Plancherel theorem we have
\begin{align}
    \label{eq:UP2}1&=\norm{\mu(H)^{-1/2}1_H}_{L^2(G_c)}^2=\norm{\mc{F}_{G_c}(\mu(H)^{-1/2}1_H)}_{L^2(\wh{G}_c)}^2\\
    \notag &=\norm{\mu(H)^{1/2}1_{H^\perp}}_{L^2(\wh{G}_c)}^2=\mu(H)\mu_{\wh{G}_c}(H^\perp)=c^{-1}\mu(H)\mu(H^\Delta).
\end{align}
It remains to show that $c=1$. Since $G_c$ admits a compact open 2-regular subgroup, so too does $F_c$ (project onto first coordinate). Let $K\leq F_c$ be such a subgroup. Then as shown in Example \ref{ex:UP}, the characteristic function of the state $\psi=\mu_{F_c}(K)^{-1/2}1_K\in L^2(F_c)$ is $\chi_{|\psi\ra\la \psi|}=1_{K\times K^\perp}$. It is easy to see that $K\times K^\perp$ is a compact open Lagrangian subgroup of $G_c$. Hence, \eqref{eq:UP2} implies
    $$\mu(K\times K^\perp)^2=\mu(K\times K^\perp)\mu((K\times K^\perp)^\Delta)=c.$$
Theorem \ref{thm-twistedPlancherel} then shows
    \begin{equation}\label{eq-KKperp}
    1=\norm{|\psi\ra\la\psi|}^2_2=\norm{1_{K\times K^\perp}}_{L^2(G_c)}^2=\mu(K\times K^\perp)=\sqrt{c}.    
    \end{equation}
\end{proof}

\begin{lem}\label{lem-Lagrangian}
Let $H$ be a compact open isotropic subgroup of $G_c$. Then $H$ is Lagrangian if and only if $\mu_{G_c}(H)=1$.
\end{lem}

\begin{proof}
Let $\mu = \mu_{G_c}$ for simplicity. If $H=H^\Delta$, then $\mu(H)=1$ is direct from Lemma \ref{lem-UP}. Conversely, if $\mu(H)=1$, then we get $\mu(H)=\mu(H^\Delta)=1$ from the conditions $\mu(H)\leq\mu(H^\Delta)$ and $\mu(H)\mu(H^\Delta) = 1$. Since $H\subseteq H^\Delta$, this implies that $H=H^\Delta$. Note that $H^\Delta$ is compact open by Lemma \ref{lem-compactopen}

\end{proof}

Combining \eqref{eq-entropyformula} and Lemma \ref{lem-Lagrangian}, together with the fact that a state is pure if and only if its entropy is 0, we get the following conclusion.

\begin{prop} \label{prop-finGauss-pure1} A B-Gaussian state $\rho_{H,\gamma}$ is pure if and only if $H$ is Lagrangian.
\end{prop}

We now show that pure B-Gaussian states in $\D(L^2(F_c))$ are determined by a point in the phase space $G_c$ and a symmetric bicharacter, which is the analogue of the first and second moments for pure bosonic Gaussian states. Recall that a bicharacter $\beta:K\times K\to \tor$ on an LCA group $K$ is {\em symmetric} if $\beta(x,y)=\beta(y,x),\; x,y\in K$.

\begin{thm} \label{thm-finGauss-pure2}
A pure state $\rho=|\psi\ra\la\psi|\in \mc{D}(L^2(F_c))$ is B-Gaussian if and only if there exists a compact open 2-regular subgroup $K$ of $F_c$, a symmetric bicharacter $\beta:K\times K\to \tor$, $z_0\in G_c$, and $\alpha\in \tor$ such that $\psi=\alpha W(z_0)\psi_0$, where
\begin{equation} \label{eq-pure-Gaussian}
    \psi_0(x)=\mu_{F_c}(K)^{-1/2} 1_{K}(x)\beta(x,2^{-1}x),\;\; x\in F_c.
\end{equation}
Alternatively, for $z_0 = (x_0,\gamma_0)$ we have
    $$\psi(x)=\tilde{\alpha}\mu_{F_c}(K)^{-1/2} 1_{K+x_0}(x)\gamma_0(x)\beta(x,2^{-1}x),\;\; x\in F_c,$$
for some $\tilde{\alpha}\in \tor$. In this case, $\rho=\rho_{H,\Gamma}$ where $\Gamma=\Delta(z_0,\cdot)$ and
\begin{equation} \label{eq-Lagrangian}
    H=\set{(x,\gamma)\in G_c:x\in K, \gamma|_K=\beta(x,\cdot)}.
\end{equation}
\end{thm}

\begin{rem}
%\begin{enumerate}
    %\item ({\bf REWRITE}) Although the characterization Theorem \ref{t:finGauss} might not be true for general $F$, every $\rho=\rho_{H,\Gamma}$ is still a B-Gaussian state for general $F$, so the above theorem gives the characterization of pure B-Gaussian states of the form $\rho_{H,\Gamma}$. If Theorem \ref{t:finGauss} is true, then the result can be easily extended to general 2-regular LCA group $F=\Real^n\times F_c$.

%\item

Note that the above theorem implies that we can choose a continuous wave function for every pure B-Gaussian state. Furthermore, the result includes the characterization of stabilizer states in \cite[Lemma 18]{Gro}, which is equivalent to saying that if $d\geq 3$ is an odd integer, every pure B-Gaussian state $\rho=|\psi\ra\la\psi|\in \D(\ell^2(\z_d^n))$ with $\psi(x)\neq 0$ for all $x\in \z_d^n$ is exactly of the form
    $$\psi(x)=d^{-n/2}\om^{x^TAx+b^T x+c},$$
where $\om=\exp(\frac{2\pi i}{d})$, $A\in M_n(\z_d)$ is a symmetric matrix, $b\in \z_d^n$, and $c\in \Real$.
%\end{enumerate}

\end{rem}

The proof of Theorem \ref{thm-finGauss-pure2} begins with a connection between compact open 2-regular Lagrangian subgroups and symmetric bicharacters, as follows.

\begin{lem} \label{lem-Lag-symmbichar}
There exists one-to-one correspondence between the family of 2-regular compact open Lagrangian subgroups $H$ of $G_c$ and the family of pairs $(K,\beta)$ consisting of a compact open 2-regular subgroup $K$ of $F_c$ and a symmetric bicharacter $\beta:K\times K\to \tor$, related by equation \eqref{eq-Lagrangian}.
\end{lem}
\begin{proof}
For a given pair $(K,\beta)$, let $H\subset G_c$ be defined by the relation \eqref{eq-Lagrangian}, which can easily be checked to be 2-regular closed subgroup of $G_c$.
The isotropy of $H$ follows from the symmetry of $\beta$: for $(x,\gamma), (x',\gamma')\in H$,
    $$\Delta((x,\gamma),(x',\gamma'))=\beta(x,x')\overline{\beta(x',x)}=1.$$
Moreover, for each $x\in K$, the corresponding section 
    \begin{equation}\label{eq-Hsection}
    H_x:=\set{\gamma\in \widehat{F}_c:(x,\gamma)\in H}=\set{\gamma\in \widehat{F}_c:\gamma|_K=\beta(x,\cdot)}    
    \end{equation}
is actually a coset of $K^{\perp}$ in $\widehat{F}_c$. By Fubini's theorem and \eqref{eq-KKperp}, we have
    $$\mu_{G_c}(H)=\int_K \mu_{\widehat{F}_c}(H_x) \,d\mu_{F_c}(x)=\int_K \mu_{\widehat{F}_c}(K^{\perp})\,d\mu_{F_c}(x)=\mu_{F_c}(K)\mu_{\widehat{F}_c}(K^{\perp})=1.$$
This implies that $H$ is open and compact by \eqref{eq-cpt-open-Haar}, and $H$ is Lagrangian by Lemma \ref{lem-Lagrangian}, which explains one direction of the correspondence.

For the reverse direction, let $H$ be a compact open 2-regular Lagrangian subgroup of $G_c$. For the natural projection $\pi_{F_c}:(x,\gamma)\in G_c\mapsto x\in F_c$, define $K:=\pi_{F_c}(H)$. Then $K$ is a 2-regular compact open subgroup of $F_c$ since $\pi_{F_c}$ is a continuous homomorphism and an open map. We first claim that for each $x\in K$, we have $H_x = \gamma_x + K^\perp$ for some $\gamma_x\in \widehat{F}_c$, where $H_x$ is from \eqref{eq-Hsection}. Indeed, we can pick any $\gamma_x\in \wh{F}_c$ such that $(x,\gamma_x)\in H$, and then
\begin{align*}
    \gamma\in H_x &\Longleftrightarrow (0, \gamma-{\gamma_x})\in H=H^{\Delta}\\
    &\Longleftrightarrow \Delta((0,\gamma-{\gamma_x}),(x',\gamma'))=\gamma(x')\overline{\gamma_x(x')}=1 \text{ for all } (x',\gamma')\in H\\
    &\Longleftrightarrow \gamma-{\gamma_x}\in K^{\perp} \Longleftrightarrow \gamma\in \gamma_x+ K^{\perp}.
\end{align*}
Next, we claim that the map $T: K \to \widehat{F}/K^{\perp},\; x \mapsto \gamma_x + K^\perp$ is a continuous homomorphism. The additivity is clear from the definition, and the continuity comes from the facts that $\widehat{F}/K^{\perp}$ is discrete and 
\begin{align*}
    \text{Ker}\, T &=\set{x\in K: (x,\gamma)\in H \text{ for some } \gamma\in K^{\perp}}\\
    &=\bigcup_{\gamma\in K^{\perp}}\set{x\in K:1_H(x,\gamma)\neq 0}
\end{align*}
is open, which in turn comes from the continuity of the function $1_H$. Passing through the canonical identification $\widehat{F}/K^{\perp}\cong \widehat{K}$ we obtain a continuous homomorphism $\tilde{T}: K \to \wh{K}$, and we can readily check that the associated bicharacter $\beta:K\times K\to \tor,\; (x,y)\mapsto \gamma_x(y)$ is the one we were looking for. Indeed, $\beta$ is symmetric since $H$ is isotropic:
    $$\beta(x,y)\overline{\beta(y,x)}=\gamma_x(y)\overline{\gamma_y(x)}=\Delta((x,\gamma_x),(y,\gamma_y))=1,\;\; x,y\in K.$$
The relation \eqref{eq-Lagrangian} is now straightforward.

Finally, one can easily check that the maps $(K,\beta)\mapsto H$ and $H \mapsto (K,\beta)$ are inverses to each other.
\end{proof}

\begin{proof}[Proof of Theorem \ref{thm-finGauss-pure2}]
Suppose $\rho=\rho_{H,\Gamma}$ is a pure B-Gaussian state. Then $H$ is a 2-regular Lagrangian compact open subgroup of $G_c$ by Proposition \ref{prop-finGauss-pure1}. By considering $\rho_0:=W(z_0)^*\rho W(z_0)$ for $z_0\in G_c$ such that $\Gamma=\Delta(z_0,\cdot)$, we may assume that $\Gamma\equiv 1$. Moreover, we can choose a pair $(K,\beta)$ as in Lemma \ref{lem-Lag-symmbichar} such that equation \eqref{eq-Lagrangian} holds.
For the conclusion we only need to check that $\chi_{\psi} = \chi_\rho$ for $\psi(x) = \mu_{F_c}(K)^{-1/2}1_K(x)\beta(x,2^{-1}x)$, $x\in F_c$.
First, we recall that $\chi_\rho(x,\gamma)=1_H(x,\gamma)=1_K(x)1_{H_x}(\gamma)$, $(x,\gamma) \in G_c$. Moreover, there is $\gamma_x \in \wh{F}_c$ such that $H_x = \gamma_x+K^{\perp}$ for each $x\in K$ as in the proof of Lemma \ref{lem-Lag-symmbichar}. Recall also that $\beta(x,y) = \gamma_x(y)$ with the above choice.
Now we observe that
    \begin{align*}
        \chi_{\psi}(x,\gamma)
        & = \int_{F_c} \overline{\gamma(y - 2^{-1}x)}\,\overline{\psi(y-x)}\psi(y)d\mu_{F_c}(y)\\
        %& = \int_{F_c} \overline{\gamma(y - 2^{-1}x)}\mu_{F_c}(K)^{-1}1_K(y-x)1_K(y)\overline{\beta(y-x, 2^{-1}(y-x))}\beta(y, 2^{-1}y)d\mu_{F_c}(y)\\
        & = \int_K \overline{\gamma(y - 2^{-1}x)}\mu_{F_c}(K)^{-1}1_K(y-x)\beta(x, y-2^{-1}x)d\mu_{F_c}(y)\\
        & = 1_K(x)\int_K \overline{\gamma(y)}\mu_{F_c}(K)^{-1}\beta(x, y)d\mu_{F_c}(y)\\
        & = 1_K(x) \int_K \overline{\gamma(y)}\mu_{F_c}(K)^{-1}\gamma_x(y)d\mu_{F_c}(y)\\
        &= 1_K(x) 1_{\gamma_x + K^{\perp}}(\gamma|_K),
    \end{align*}
which lead to the desired conclusion.

We also get the converse by following the above calculation process backwards, again combined with Lemma \ref{lem-Lag-symmbichar}.
\end{proof}

It is well-known that in bosonic systems, every Gaussian state belongs to the norm-closed convex hull of pure Gaussian states (\cite[Problem 5.10]{Serafini}). The same phenomenon occurs in our setting.

\begin{lem}\label{lem-normconvint} Let $H$ be a compact open 2-regular isotropic subgroup of $G_c$. The map $\wh{G}_c\ni\gamma\mapsto\rho_{H,\gamma}\in\mc{S}^1(L^2(F_c))$ is norm continuous.
\end{lem}

\begin{proof} Take a net $(\gamma_i)$ converging to $\gamma\in\wh{G}_c$, meaning uniform convergence on compact sets. Since $H$ is a compact open subgroup, it follows that $\gamma_i1_H\rightarrow\gamma1_H$ in $L^2(G_c)$. Thus, by continuity of the twisted Fourier transform (Theorem \ref{thm-twistedPlancherel})
$$\rho_{H,\gamma_i}=\F^\sigma_G(\gamma_i1_H)\rightarrow\F^\sigma_G(\gamma1_H)=\rho_{H,\gamma}$$
in $\mc{S}^2(L^2(F_c))$. Since $\rho_{H,\gamma_i}^2=\mu_{G_c}(H)\rho_{H,\gamma_i}$ (Proposition \ref{prop-fin-square}) we have $\sqrt{\rho_{H,\gamma_i}}=\mu_{G_c}(H)^{-1/2}\rho_{H,\gamma_i}$. Similarly, $\sqrt{\rho_{H,\gamma}}=\mu_{G_c}(H)^{-1/2}\rho_{H,\gamma}$. Hence, $$\norm{\sqrt{\rho_{H,\gamma_i}}-\sqrt{\rho_{H,\gamma}}}_2=\mu_{G_c}(H)^{-1/2}\norm{\rho_{H,\gamma_i}-\rho_{H,\gamma}}_2\rightarrow 0$$ 
in $\mc{S}^2(L^2(F_c))$. Furthermore, isotropy of $H$ implies $\rho_{H,\gamma_i}\rho_{H,\gamma}=\rho_{H,\gamma}\rho_{H,\gamma_i}$ and
$$\norm{\rho_{H,\gamma_i}-\rho_{H,\gamma}}_1\leq\norm{\sqrt{\rho_{H,\gamma_i}}-\sqrt{\rho_{H,\gamma}}}_2\norm{\sqrt{\rho_{H,\gamma_i}}+\sqrt{\rho_{H,\gamma}}}_2\rightarrow 0.$$
\end{proof}

\begin{thm}
Every B-Gaussian state in $\mc{S}^1(L^2(F))$ belongs to the norm closed convex hull of pure B-Gaussian states.
\end{thm}
\begin{proof}
Thanks to the decomposition $\rho_n\ten \rho_c$ we can focus on the case of the state $\rho_c = \rho_{H,\gamma}$ for some compact open 2-regular isotropic subgroup $H$ of $G_c$ and a character $\gamma \in \wh{H}$. Pick a maximal isotropic subgroup $K$ containing $H$ (by Zorn's lemma, if needed). %of $H^\Delta$ (by Zorn's lemma, if needed). 
Since $H^\perp$ is a compact open subgroup of $\wh{G}_c$ and the map $\gamma'\mapsto \rho_{K ,\gamma\gamma'}$ is continuous by Lemma \ref{lem-normconvint}, the following state is well defined.
    $$\rho = \frac{1}{\mu_{\wh{G}_c}(H^\perp)}\int_{H^\perp}\rho_{K ,\gamma\gamma'}d\mu_{\wh{G}_c}(\gamma').$$
We only need to check that $\chi_{\rho_{H,\gamma}} = \chi_\rho$ for the desired conclusion by Proposition \ref{prop-finGauss-pure1}. Indeed, for $z\in G_c$ we have
\begin{align*}
    \chi_\rho(z)&=\tr\bigg(W_{1/2}(z)^*\bigg(\frac{1}{\mu_{\wh{G}_c}(H^\perp)}\int_{H^\perp}\rho_{K ,\gamma\gamma'}d\mu_{\wh{G}_c}(\gamma')\bigg)\bigg)\\
    &=\frac{1}{\mu_{\wh{G}_c}(H^\perp)}\int_{H^\perp}\tr(W_{1/2}(z)^*\rho_{K ,\gamma\gamma'})d\mu_{\wh{G}_c}(\gamma')\\
    & = \frac{1}{\mu_{\wh{G}_c}(H^\perp)}\int_{H^\perp}\gamma(z)\gamma'(z) 1_{K}(z) d\mu_{\wh{G}_c}(\gamma')\\
    &=\frac{\gamma(z)1_{K}(z)}{\mu_{\wh{G}_c}(H^\perp)}\int_{H^\perp}\gamma'(z)d\mu_{\wh{G}_c}(\gamma')\\
    &=\gamma(z)1_{K }(z) 1_{H^{\perp\perp}}(z) = \gamma(z)1_{K \cap H}(z)\\
    &=\gamma(z)1_H(z) = \chi_{\rho_{H,\gamma}}(z).
\end{align*}
\end{proof}

\section{Angle-number systems}\label{sec-AN}

In this section we show that B-gaussian states in the angle-number system in $d$-modes are nothing but the pure states whose wave functions are the elements of the canonical orthonormal basis $\{ |m\ra = |e_m\ra : m\in \z^d\} \subseteq \Hi = L^2(\tor^d)\cong \ell^2(\z^d)$, where $e_m(\theta) = e^{2\pi i \la \theta, m\ra}$, $\theta \in \tor^d$. Recall that the associated Weyl representation $W_{1/2}$ is
    $$W_{1/2}(\theta,n):=e^{\pi i\la  \theta, n \ra}T_\theta M_n,\;\; (\theta,n)\in \tor^d\times \z^d.$$
See Section \ref{ss:ANS} for details.

The first step of the proof is to determine the characteristic functions for rank-1 operators acting on $\Hi$.

\begin{lem}\label{lem-char-ft-angle-number}
For $a,b\in \z^d$ we have 
\begin{equation} \label{eq-char-angle-number}
    \chi_{|a\ra \la b|}(\theta, n)=\delta_{a-b, n}\;e^{\pi i \la \theta, a+b\ra},\;\; (\theta,n)\in \tor^d\times \z^d.
\end{equation}
\end{lem}
\begin{proof}
It is straightforward from the computation
\begin{align*}
    \chi_{|a\ra\la b|}(\theta,n)&=\la b|W_{1/2}(-\theta,-n)|a\ra\\
    &=\int_{\tor^d} e^{-2\pi i \la \theta',b \ra}e^{\pi i \la \theta, n\ra} e^{2\pi i \la(\theta'+\theta),a-n\ra}\,d\theta'\\
    &=e^{\pi i \la \theta, 2a-n\ra}\int_{\tor^d} e^{2\pi i \la \theta',a-b-n \ra}\,d\theta' =\delta_{a-b,n}e^{\pi i \la \theta, a+b\ra}.
\end{align*}

\end{proof}
We again remark that the formula \eqref{eq-char-angle-number} is only valid for our identification $\theta\in \left[-\frac{1}{2},\frac{1}{2} \right)^d$ through \eqref{eq-angle-number-duality}.

\begin{thm}\label{thm-B-GaussianState-ch}
The set of all B-Gaussian states for the angle-number system in $d$-modes is the set of all pure states of the form $|m \ra \la m|$ for some $m\in \z^d$.
\end{thm}
\begin{proof}
Let $\rho$ be a B-Gaussian state with the (open) support $H$ of $\chi_\rho$. Since $H$ is an open subgroup of $G = \tor^d \times \z^d$ we know that $H = \tor^d \times K$ for a subgroup $K$ of $\z^d$. It is easy to check that the Haar measure $\mu$ on $G$ respecting the twisted Plancherel formula \eqref{eq-twisted-Plancherel} is given by
    $$\int_G f\, d\mu = \sum_{n\in \z^d}\int_{\tor^d} f(\theta, n) d\theta,\;\; f\in C_c(G).$$
Then, by \eqref{eq-twisted-Plancherel} and Lemma \ref{lem-char-ft-angle-number}, we have
\begin{align*}
    \la a|\rho|b \ra &= \tr (\rho (|a\ra\la b|)^*) = \sum_{n\in \z^d}\int_{\tor^d} \chi_{\rho}(\theta,n) \overline{\chi_{|a\ra\la b|}(\theta,n)}\,d\theta\\
    &=\sum_{n\in K}\int_{\tor^d} \delta_{a-b,n} \chi_{\rho}(\theta,n) e^{-\pi i\la\theta,a+b\ra}\,d\theta\\
    &=\int_{\tor^d} 1_K(a-b) \chi_{\rho}(\theta,a-b)e^{-\pi i\la\theta,a+b\ra}\,d\theta
\end{align*}
for $a,b\in \z^d$.
In particular,
    $$\la a|\rho|a \ra = \int_{\tor^d} \chi_\rho(\theta,0) e^{-2\pi i \la \theta,a \ra}\,d\theta = \widehat{\chi_{\rho}(\cdot,0)}(a),\; a\in \z^d.$$
On the other hand, the B-Gaussianity of $\rho$ again implies that $g(\cdot):=\chi_\rho(\cdot,0)$ is positive definite and satisfies the B-Gaussian identity  \eqref{e:BGauss} on $\tor^d$. Thus, $g$ is the Fourier transform of a B-Gaussian distribution on $\z^d\cong \widehat{\tor^d}$ by Bochner's theorem. Furthermore, we note that $\z^d$ contains no subgroup homeomorphic to $\tor^2$ and $g$ is nowhere vanishing. Then, Proposition \ref{prop-B-Gaussian}(2) tells us that $g$ is the Fourier transform of a Gaussian distribution. If we write $g(\theta)=e^{2\pi i\la \theta, m \ra}\exp(-\varphi(\theta))$ for some $m\in \z^d$ and continuous $\varphi:\tor^d\to [0,\infty)$ satisfying \eqref{e:parallel}, then compactness of $\tor^d$ and Remark \ref{rem-B-Gaussianity}(2) says that $\varphi\equiv 0$ since ${\rm Hom}(\tor^d, \Real) = \{0\}$. Consequently, we have $\la a|\rho|a\ra =\hat{g}(a)=\delta_{m,a}$.

The above computation means that the diagonal part of the operator $\rho$ (as an infinite matrix) is zero except one point. Thus, we can conclude that off-diagonal parts of the positive operator $\rho$ must be zero. This forces $\rho = |m\ra \la m|$.

\end{proof}

Recalling the fact that the Fourier transform of a Gaussian distribution has full support we get the following conclusion.

\begin{cor}\label{c:angle-number-nongauss}
    There is no Gaussian state for the angle-number system in $d$-modes.
\end{cor}

\begin{rem}
The above characterization is consistent with the results about characterizing pure states with non-negative Wigner functions on the angle-number system in 1-mode \cite{rigas2010non}.
\end{rem}

\section{Fermions and hard-core bosons}\label{sec-second-ex}

In this section we show that there are no B-Gaussian states in the fermionic and hard-core bosonic systems, introduced in Section \ref{ss:fermion}. Although stabilizer states exist and are heavily studied in these qubit systems, in comparison with our previous results on finite 2-regular groups, this section shows that qubit stabilizer states do not possess an underlying Gaussian characterization in the sense of Bernstein.

We begin with a simple description of B-Gaussian distributions on $\z_2^m$.

\begin{prop} \label{prop-qubit-Gaussian}
Every B-Gaussian distribution on $G=\z_2^{m}$ is of the form $\delta_a$ for some $a\in G$, which is a Gaussian distribution on $G$.
\end{prop}

\begin{proof}
    Let $\mu$ be a B-Gaussian distribution on $G=\z_2^{m}$ and let $H={\rm supp}\,\hat{\mu}$. Then the annihilator $H^\perp$ is trivial (or equivalently, $H=G$) since it is a compact Corwin subgroup of $G$ in which all elements have order 2. Thus, Proposition \ref{prop-B-Gaussian} (2) tells us that $\mu$ is a Gaussian distribution on $G$. In particular, $\hat{\mu}$ is a character on $\widehat{G}$ as the associated quadratic function $\varphi$ must vanish, which means that $\mu$ is a point-mass at some point on $G$. Note finally that it is straightforward to see that every point-mass is a Gaussian distribution.

\end{proof}

Let us first focus on the hard-core boson setting.

\begin{thm} \label{thm-hardcore-nonGauss}
   For any choice of normalizing factor $\xi$ \eqref{eq-normalization-hardcore}, there is no B-Gaussian state on the quantum kinematical system $(\z_2^n\times \z_2^n, \tilde{\sigma}_{\rm can})$.
\end{thm}

\begin{proof}
    %\tcr{(Our original proof was based on a specific choice of  $\xi$, which featured the factoring property. By the same idea with fermionic case, we can write general statement for any choice of $\xi$)}
    Suppose $\rho\in \D((\Comp^2)^{\otimes n})$ is a B-Gaussian state associated to the (B-)Gaussian distribution $\delta_a$, $a\in \z_2^{2n}$ (Proposition \ref{prop-qubit-Gaussian}). By Equation \eqref{eq-chftnclifford} and the non-degeneracy of $\Delta$, there exists $z_0\in \z_2^{2n}$ such that $\rho_0:=W(z_0)^*\rho W(z_0)$ has a characteristic function
        $$\chi_{\rho_0}(w)=\overline{\Delta(z_0,w)}\widehat{\delta_a}(w)=(-1)^{z_0^T J w}(-1)^{a^T w}\equiv 1.$$
    However, the twisted Fourier inversion (Proposition \ref{prop-Fourier-char}) gives that
        $$\rho_0=\frac{1}{2^n}\sum_{z\in \z_2^{2n}} W_{1/2, \rm can}(z)=\frac{1}{2^n}\sum_{z\in \z_2^{2n}} \xi(z)W_{\rm can}(z),$$
    and the RHS must define a state. On the other hand, for $z=(x_1,\ldots, x_n, y_1,\ldots, y_n)\in \z_2^{2n}$, observe from \eqref{eq-W-hardcore} that
    $$(\id\otimes\cdots\otimes\id\otimes\tr)W_{\rm can}(z)=2\delta_{0,x_{n}}\delta_{0,y_{n}} h^{x_1}_1 h^{y_1}_2\cdots h^{x_{n-1}}_{2n-3}h^{y_{n-1}}_{2n-2}.$$
    By repeating the procedure, we get
        $$(\id\otimes\tr\otimes\cdots\otimes\tr)W_{\rm can}(z )=2^{n-1}\delta_{0,x_{2}}\delta_{0,y_{2}}\cdots\delta_{0,x_{n}}\delta_{0,y_{n}} h^{x_1}_1 h^{y_2}_2.$$
    Therefore,
    \begin{align*}
        (\id\otimes\tr\otimes\cdots\otimes\tr)\rho_0&=\frac{1}{2}\sum_{x_1,y_1\in \z_2} \xi(x_1e_1,y_1e_1)h^{x_1}_1h^{y_1}_2\\
            &=\frac{1}{2}(I\pm X\pm Y\pm Z)
    \end{align*}
    where $e_1=(1,\ldots, 0)\in \z_2^n$, from the formulae $\xi(e_1,0)^2=\xi(0, e_1)^2=1$ and $\xi(e_1,e_1)^2=-1$. But it is easy to see that, for any choice of signs, the resulting operator is not positive, a contradiction.
    %By ?? and the unitary equivalence of B-Gaussian states, it suffices to show that there is no state $\rho$ with $\chi_{\rho}\equiv 1$. In this case, the factoring property of $W_{1/2}$,
    %\begin{equation}
    %    W_{1/2}(x_1,\ldots,x_d,y_1,\ldots,y_d)=\bigotimes_{i=1}^d \xi_1(x_i,y_i)X^{x_i}Z^{y_i},\;\; x_i,y_i\in \z_2,
    %\end{equation}
    %implies that
    %\begin{align*}
    %    \sum_{ a \in \z_2^d}W_{1/2}( a )&=\bigotimes_{i=1}^d\sum_{x_i,y_i\in \z_2} \xi_1(x_i,y_i)X^{x_i}Z^{y_i} \\
    %        &=(I+X+Y+Z)^{\otimes d}=\begin{bmatrix} 2 & 1-i \\ 1+i & 0 \end{bmatrix} ^{\otimes d}.
    %\end{align*}
    %Therefore, if $\chi_{\rho}\equiv 1$, then $\rho=\frac{1}{2^d}\sum_{ a \in \z_2^d}W_{1/2}( a )$ cannot be positive, a contradiction.
\end{proof}

The same method works for fermionic systems.

%Let us turn our attention to the case of fermionic systems. 

\begin{thm} \label{thm-fermi-nonGauss}
   For any choice of the normalizing factor $\xi$ \eqref{eq-normalization-fermi}, there is no B-Gaussian state on the quantum kinematical system $(\z_2^{2n}, \tilde{\sigma}_{\rm fer})$.
\end{thm}
\begin{proof}
As in the hardcore boson case it boils down to check the operator
    $$\rho=\frac{1}{2^n}\sum_{ a \in \z_2^{2n}}\xi( a ) W_{\rm fer}( a )$$
is not positive.
%From the definition \eqref{eq-W-eps}, we observe that $$(\id\otimes\cdots\otimes\id\otimes\tr)W_{\rm fer}( a )=2\delta_{0,x_{2d-1}}\delta_{0,x_{2d}} c^{x_1}_1 \cdots c^{x_{2d-2}}_{2d-2}.$$
%    By repeating the procedure, we get
By the same argument in Theorem \ref{thm-hardcore-nonGauss}, we have
$$(\id\otimes\tr\otimes\cdots\otimes\tr)W_{\rm fer}(a)=2^{n-1}\delta_{0,x_{3}}\delta_{0,x_4}\cdots\delta_{0,x_{2n}} c^{x_1}_1c^{x_2}_2$$
for $a=(x_1,\ldots, x_{2n})\in \z_2^{2n}$, and therefore,
\begin{align*}
    (\id\otimes\tr\otimes\cdots\otimes\tr)\rho&=\frac{1}{2}\sum_{x_1,x_2\in \z_2}\xi(x_1,x_2,0,\ldots,0)c^{x_1}_1c^{x_2}_2\\
    &=\frac{1}{2}(I\pm X\pm Y\pm Z),
\end{align*}
    %from the formulae $\xi(1,0,0,\ldots, 0)^2=\xi(0,1,0,\ldots,0)^2=1$ and $\xi(1,1,0,\ldots,0)^2=-1$. But it is easy to see that, for no choice of the signs, the resulting partial trace of the operator is not positive.
which is a contradiction as before.
\end{proof}

\section{Hudson's Theorem for 2-regular totally disconnected groups} \label{sec-Hudson}

\textit{Hudson's theorem}~\cite{hudson1974wigner} and its higher dimensional generalization~\cite{soto1983wigner} show that %among pure states, it is exactly the class of Gaussian states.
pure bosonic Gaussian states can be characterized by non-negativity of their Wigner functions. Gross \cite{Gro} continued this line of research for the Weyl system with $F=\z_d^n$, $d(\geq 3)$ odd, characterizing pure states with non-negative Wigner functions as the class of stabilizer states, i.e. pure B-Gaussian states in our terminology. We extend the result of Gross to the case of totally disconnected groups. Recall that a topological space is \textit{totally disconnected} if the only connected sets are singletons. Note that our proof is inspired by the one of Gross \cite{Gro}, but there are fundamentally new aspects to accommodate the infinite group setting.

In this section, $F$ denotes a (second countable) 2-regular totally disconnected LCA group, unless otherwise noted.

\begin{prop}\label{prop-Dantzig}(\textit{van Dantzig}, \cite{Da36}, \cite[Theorem 7.7]{HewittRoss})
Every open neighborhood of the identity of a totally disconnected locally compact
group contains a compact open subgroup.
\end{prop}

Since $F$ contains a compact open subgroup, all the facts from Section \ref{sec-compactopen} are applicable to the kinematical system $(G=F\times \widehat{F}, \sigma=\tilde{\sigma}_{\rm can})$ with the corresponding Weyl representation $W=W_{1/2}$ given by
    $$W(x,\gamma)\psi(y)= \overline{\la 2^{-1}x,\gamma \ra} \la y, \gamma\ra \psi(y-x),\;\; \psi\in L^2(F), \; x, y\in F, \gamma \in \wh{F}.$$
Let us express the Wigner function $\mathcal{W}_{\psi}$ of a vector state $\psi\in L^2(F)$ using {\em the self-correlation function} as in \cite[p.10]{Gro},
    $$\varphi_q(x):=\psi(q+2^{-1}x)\overline{\psi(q-2^{-1}x)},\;\; q,x\in F.$$
We first note that
    $$\chi_{\psi}(x,\gamma)=\int \la 2^{-1}x,\gamma\ra \overline{\la y,\gamma \ra}\,\overline{\psi(y-x)}\psi(y)d\mu_F(y)=\la 2^{-1}x,\gamma \ra \wh{g_x}^F(\gamma)$$
with $g_x(y)=\overline{\psi(y-x)}\psi(y)$, $x, y\in F, \gamma \in \wh{F}$. On the other hand we have
    $$\Delta((q,p),(x,\gamma)) = p(x)\overline{\gamma(q)},\;\; q,x\in F, p,\gamma \in \wh{F}.$$
Combining the above we get
\begin{align} \label{eq-totdisconn-Wigner}
    \mathcal{W}_{\psi}(q,p)&=\left[\mathcal{F}^F\otimes (\mathcal{F}^{F})^{-1}\right](\chi_{\psi})(p,q)
    =(\mathcal{F}^F g_{\cdot}(q+2^{-1}\cdot))(p)\nonumber\\
    %=\int_F g_x(q+2^{-1}x)\overline{\la x, p\ra}\,d\mu_F(x)\nonumber\\
    %&=\int_F \overline{\psi(q-2^{-1}x)}\psi(q+2^{-1}x)\overline{\la y, p\ra}\,d\mu_F(y)\nonumber\\
    & = \wh{\varphi_q}^F(p),\;\; q\in F, p \in \wh{F}.
\end{align}

The main theorem of this section is the following.

\begin{thm}[\textbf{Hudson's theorem, 2-regular totally disconnected version}]\label{thm-Hudson-totdisconn}
For a pure state $\psi\in L^2(F)$ over the Weyl system $(F\times \widehat{F}, \tilde{\sigma}_{\rm can})$, the following are equivalent:
\begin{enumerate}
    \item $\rho=|\psi\ra\la \psi|$ is B-Gaussian,
    
    \item $\psi$ is continuous and $\mathcal{W}_{\psi}\geq 0$ a.e.
\end{enumerate}
\end{thm}

%Note that Theorem \ref{thm-Hudson-totdisconn} includes the cases where $F$ is a 2-regular finite group or $F=\bQ_p^n$ with odd prime $p$. In particular, the case $F=\z_d^n$ with odd $d$ recovers the discrete Hudson theorem by D. Gross \cite{Gro}.

The proof for the direction $(1)\Rightarrow(2)$ is a simple combination of Theorem \ref{t:finGauss} and Theorem \ref{thm-finGauss-pure2}. Indeed, a B-Gaussian pure state $\rho=\rho_{H,\Gamma}$ associated to a Lagrangian subgroup $H$ and a character $\Gamma=\Delta(z_0,\cdot)$ has a characteristic function $\chi_{\rho}=\Gamma\cdot  1_H$. Therefore we have $\W_{\psi}(z)=1_H(z-z_0)\geq 0$. Moreover, \eqref{eq-pure-Gaussian} reveals that $\psi$ is continuous.

The reverse direction $(2)\Rightarrow(1)$ is the main difficulty. Let us begin with a lemma which exploits the total disconnectedness of $F$ in a crucial way.

\begin{lem} \label{lem-totdisconn-integrable}
If $f\in L^1(F)$, $\hat{f}\geq 0$ a.e., and if $f$ is continuous at $0$, then $\hat{f}\in L^1$.
\end{lem}
\begin{proof}
Proposition \ref{prop-Dantzig} and second countability of $F$ give a sequence $\set{K_n}_{n=1}^{\infty}$ of compact open subgroups of $F$ decreasing to the trivial subgroup. Now we claim that $1_{(K_n)^{\perp}}\to 1$ pointwise on $\wh{F}$ as $n\to \infty$. Indeed, if $\gamma\in \wh{F}$ and $\epsilon\in (0,\frac{1}{2})$, then $V=\set{x\in F:\, |\la x,\gamma\ra-1|<\epsilon}$ is a neighborhood of $0$. Choose $N$ such that $K_N\subset V$. Since $K_n\subset V$ for $n\geq N$, we have
\begin{align*}
    |1_{(K_n)^{\perp}}(\gamma)-1|&=\left|\mu_F(K_n)^{-1}\,\widehat{1_{K_n}}(\gamma)-1\right|\\
    &=\left|\int_V (\overline{\la x,\gamma\ra}-1)\mu_F(K_n)^{-1}\,1_{K_n}(x)\,dx\right|<\epsilon\,(<1/2).
\end{align*}
Since $|1_{(K_n)^{\perp}}(\gamma)-1|$ is either 0 or 1, we have $1_{(K_n)^{\perp}}(\gamma)=1$ for all $n\geq N$.

Now we apply the monotone convergence theorem and Fubini's theorem together with the above claim to get
\begin{align*}
    \int_{\widehat{F}} \hat{f}(\gamma)d\gamma &=\lim_{n\to \infty}\int_{\widehat{F}}\hat{f}(\gamma)1_{(K_n)^{\perp}}(\gamma)d\gamma\\
    &=\lim_{n\to \infty}\int_{F} \int_{\widehat{F}} f(x)1_{(K_n)^{\perp}}(\gamma)\overline{\la x,\gamma\ra}\,d\gamma\,dx\\
    %&=\lim_{n\to \infty}\int_{F} f(x)(1_{(K_n)^{\perp}}){^{\vee}}(-x)dx\\
    &=\lim_{n\to \infty}\mu_F(K_n)^{-1}\int_{F} f(x){1_{K_n}}(x)dx\\
    &=f(0)<\infty.
\end{align*}
Note that we used the continuity of $f$ at $0$ for the last equality.
\end{proof}

We proceed with an analogue of \cite[Lemma 11]{Gro}.

\begin{lem} \label{lem-totdisconn-Hudson1} If $\psi \in L^2(F)$ is continuous and $\mathcal{W}_{\psi}\geq 0$ a.e., then
$\varphi_q$ is a positive definite function on $F$ for each $q \in F$. Moreover, we have
\begin{equation}\label{eq-totdisconn-modulus1}
    |\psi(q)|^2\geq |\psi(q+x)|\,|\psi(q-x)|,
\end{equation}
\begin{equation}\label{eq-totdisconn-modulus11}
    |\psi(2^{-1}(x+y))|^2\geq |\psi(x)|\,|\psi(y)|,
\end{equation}
and
\begin{equation}\label{eq-totdisconn-modulus2}
    |\varphi_q(2^{-1}(x+y))|^2\geq |\varphi_q(x)|\,|\varphi_q(y)|
\end{equation}
for all $q,x,y\in F$.
\end{lem}
\begin{proof}
Since $\psi$ is continuous, we know that $\varphi_q$ is also continuous for all $q\in F$. From our assumption and \eqref{eq-totdisconn-Wigner} we have $\widehat{\varphi_q}^F = \mathcal{W}_{\psi}(q,\cdot) \geq 0$ a.e.. Moreover, we know $\varphi_q\in L^1(F)$ since $\psi(q\pm 2^{-1}\cdot)\in L^2(F)$, so we can appeal to Lemma \ref{lem-totdisconn-integrable} to conclude that $\widehat{\varphi_q}^F$ is integrable. This implies  that $\varphi_q$ is positive definite on $F$ for all $q\in F$ from Fourier inversion.

Now, the positivity of the matrix $\footnotesize\begin{bmatrix} \varphi_q(0) & \varphi_q(2x) \\ \varphi_q(-2x) & \varphi_q(0) \end{bmatrix}$ gives
    $$\varphi_q(0)^2-\varphi_q(2x)\varphi_q(-2x)=|\psi(q)|^4-|\psi(q+x)|^2|\psi(q-x)|^2\geq 0,\; x\in F,$$
which is \eqref{eq-totdisconn-modulus1}. It is easy to see that \eqref{eq-totdisconn-modulus1} and \eqref{eq-totdisconn-modulus11} are equivalent thanks to 2-regularity and we can apply the latter to get
\begin{align*}
    |\varphi_q(2^{-1}(x+y))|^2&=|\psi(q+2^{-2}(x+y))|^2\times |\psi(q-2^{-2}(x+y))|^2\\
    &\geq |\psi(q+2^{-1}x)||\psi(q+2^{-1}y)|\times|\psi(q-2^{-1}x)||\psi(q-2^{-1}y)|\\ &=|\varphi_q(x)|\,|\varphi_q(y)|,\;\;q,x,y\in F.
\end{align*}
\end{proof}

The above lemma has an immediate consequence, which will be crucial for the proof of the main theorem.

\begin{cor} \label{cor-totdisconn-supp}
Suppose $\psi \in L^2(F)$ is continuous and non-zero with $\mathcal{W}_{\psi}\geq 0$.
The set ${\rm supp}\,\psi$ is balanced (i.e. $x,y\in {\rm supp}\,\psi$ implies $2^{-1}(x+y) \in {\rm supp}\,\psi$) and contains a coset of a compact open subgroup of $F$. Moreover, $|\psi|$ is constant on any such coset.
\end{cor}
\begin{proof}
The set ${\rm supp}\,\psi$ is obviously balanced from the inequality \eqref{eq-totdisconn-modulus11}. Since $\psi$ is continuous and not identically zero, ${\rm supp}\,\psi$ is a nonempty open set and the second assertion follows by Proposition \ref{prop-Dantzig}. For the last statement we consider a compact open subgroup $K$ of $F$ and $x\in F$ with $x+K\subseteq\mathrm{supp}\,\psi$. The function $|\psi|$ achieves a minimum $m_x>0$ on $x+K$, say at $x_m$, by continuity. However, \eqref{eq-totdisconn-modulus1} implies that
    $$m_x^2=|\psi(x_m)|^2\geq |\psi(x_m+y)||\psi(x_m-y)|\geq m_x^2,\;\;y\in K,$$
which forces $|\psi(x_m+y)|=|\psi(x_m-y)|=m_x$ for all $y\in K$. Since $x_m+K=x+K$, this means that $|\psi|\equiv m_x$ on $x+K$.
\end{proof}

The next is the most important step towards the proof of Theorem \ref{thm-Hudson-totdisconn}. It says that the function $|\psi|$ is constant on its support, which happens to be a coset of a compact open 2-regular subgroup of $F$.

\begin{lem} \label{lem-totdisconn-Hudson2}
If $\psi \in L^2(F)$ is a continuous state and $\mathcal{W}_{\psi}\geq 0$, then there exist $x_0\in F$ and a compact open 2-regular subgroup $K$ of $F$ such that
\begin{equation}\label{eq-psi}
    |\psi|=\mu_F(K)^{-1/2}1_{x_0+K}
\end{equation}
\end{lem}

For the proof of Lemma \ref{lem-totdisconn-Hudson2} we consider the following subsets of $F$:
\begin{align}\label{eq-def-Lq}
    K_q&:=\set{x\in F: |\varphi_q(x)|=\varphi_q(0)=|\psi(q)|^2}, \nonumber\\
    K_q^{\epsilon}&:=\set{x\in F: |\varphi_q(x)|\geq \epsilon},\nonumber\\
    L_q&:=\set{x\in F: |\varphi_q(x)|>0}=\set{x\in F:q\pm 2^{-1}x\in {\rm supp}\,\psi},
\end{align}
for continuous $\psi \in L^2(F)$ with $\mathcal{W}_{\psi}\geq 0$, $q\in F$ and $\epsilon>0$. It is obvious that
    $$K_q\subset K_q^{\epsilon}\subset L_q=\bigcup_{\epsilon>0}K_q^{\epsilon}$$
for $q\in {\rm supp}\,\psi$ and $0<\epsilon< |\psi(q)|^2$. The following lemma shows that the three sets are actually identical.

\begin{lem} \label{lem-totdisconn-threesets}
If $q\in {\rm supp}\,\psi$ for continuous $\psi \in L^2(F)$ with $\mathcal{W}_{\psi}\geq 0$, then $K_q$ is a 2-regular compact open subgroup of $F$, and $K_q=K_q^{\epsilon}=L_q\,$ for $\,0<\epsilon< |\psi(q)|^2$.
\end{lem}
\begin{proof}
We first check that $K_q$ is a compact open subgroup. Proposition \ref{prop-posdef} says that $K_q$ is a closed subgroup. Since ${\rm supp}\,\psi$ is an open set containing $q$ there is a compact open subgroup $K$ such that $q+K\subset {\rm supp}\,\psi$ by Proposition \ref{prop-Dantzig}. Then $2K\subset K_q$ by the fact that $|\psi| \equiv |\psi(q)|$ on $q+K$ (Corollary \ref{cor-totdisconn-supp}) and by the definition of $\varphi_q$. Since $2K$ is open ($F$ being 2-regular), $K_q$ has nonempty interior, and is therefore clopen. Moreover, as $\varphi_q\in L^1(F)$, we have
    $$\mu_F(K_q)\varphi_q(0)=\int_{K_q}|\varphi_q(x)|dx\leq \|\varphi_q\|_{L^1(F)}<\infty.$$
Consequently, $\mu_F(K_q) <\infty$, which means $K_q$ is compact.

%By Proposition \ref{prop-posdef} and Corollary \ref{cor-totdisconn-supp}, $K_q$ is a closed subgroup of $F$, and $|\varphi_q|$ is constant on each coset of $K_q$. 

Let us move our attention to $K_q^{\epsilon}$, $\epsilon\in (0,|\psi(q)|^2)$, a nonempty closed subset of $F$. By Proposition \ref{prop-posdef}(2), $|\vphi_q|$ is constant on the cosets of $K_q$, so that
    \begin{equation}\label{eq-KqKqepsilon}
        x+K_q\subset K_q^{\epsilon}\;\;\text{for any}\; x\in K_q^{\epsilon}.
    \end{equation}
Thus, $K_q^{\epsilon}$ is a union of cosets of $K_q$, and in particular, is open. Moreover, we can observe that $K_q^{\epsilon}$ is actually a finite union of cosets of $K_q$, i.e.
\begin{equation}\label{eq-Kq}
    K_q^{\epsilon}=\bigcup_{i=1}^n(x_i+K_q),\;\; x_i \in K_q^\epsilon,\; 1\le i\le n.
\end{equation}
Indeed, we have
    $$\mu_F(K_q^{\epsilon})\leq \epsilon^{-1}\int _{K_q^{\epsilon}}|\varphi_q(x)|dx\leq \epsilon^{-1}\|\varphi_q\|_{L^1(F)}<\infty,$$
which gives us the observation since cosets are disjoint with the same (non-zero) Haar measure as $K_q$.

Now let us show that $K_q^{\epsilon}$ is a subgroup of $F$. The fact that $K_q^{\epsilon}$ is closed under the inversion $x\mapsto -x$ comes from $|\varphi_q(x)|=|\varphi_q(-x)|$, $x\in F$. In order to show $K_q^{\epsilon}$ is closed under addition, we first observe that $K_q^{\epsilon}$ is closed under the map $x\mapsto 2^{-1}x$ by \eqref{eq-totdisconn-modulus2} with $y=0$. Thus, it suffices to show that $2K_q^{\epsilon}\subset K_q^{\epsilon}$ from the identity $x+y=2^{-1}(2x+2y)$. To this end, we only need to check that $2x_k\in K_q^{\epsilon}$ for $1\le k\le n$.
We will focus on the case of $x_1$ for simplicity.
Since $K_q^{\epsilon}$ is closed under the map $x\mapsto 2^{-1}x$ we get a sequence $\set{2^{-j}x_1}_{j=1}^{\infty}$ in $K_q^{\epsilon}$. From \eqref{eq-Kq} we can pick $1\le i\le n$ and $0\leq j_1<j_2$ such that $2^{-j_l}x_1\in x_i+K_q$, $l=1,2$. In particular, there exist $y_1, y_2\in K_q$ such that $2^{-j_l}x_1=x_i + y_l$, $l=1,2$. But then, as $j_2\geq j_1+1$,
    $$2^{j_2-j_1}x_1=2^{j_2}x_i+2^{j_2}y_1=(x_1-2^{j_2}y_2)+2^{j_2}y_1.$$
Therefore,
    $$2x_1=2^{-(j_2-j_1-1)}(x_1-2^{j_2}(y_2-y_1))\in K_q^{\epsilon},$$
since $x_1-2^{j_2}(y_2-y_1)\in x_1+K_q\subset K_q^{\epsilon}$ and $K_q^{\epsilon}$ is closed under the map $x\mapsto 2^{-1}x$.

So far, we have shown that $K_q^{\epsilon}$ is 2-regular compact open subgroup of $F$. Note that we have $q+2^{-1}K_q^{\epsilon}=q+K_q^{\epsilon}\subset {\rm supp}\,\psi$ from the definition of $K_q^{\epsilon}$ and $\varphi_q$, which allows us to use Corollary \ref{cor-totdisconn-supp} to get $|\psi|\equiv |\psi(q)|$ on $q+K_q^{\epsilon}$. Now it follows that $K_q^{\epsilon}\subset  K_q$, and hence $K_q^{\epsilon}=K_q$.

Finally, $L_q=\bigcup_{\epsilon>0}K_q^{\epsilon}=K_q$.
\end{proof}

Now we are ready to go back to the proof of Lemma \ref{lem-totdisconn-Hudson2}.

\begin{proof}[Proof of Lemma \ref{lem-totdisconn-Hudson2}]
We may assume $0\in {\rm supp}\,\psi$ by considering $\psi_0=W(x_0,0)^*\psi=\psi(\cdot+x_0)$ for any chosen $x_0\in {\rm supp}\,\psi$ if necessary. We claim that ${\rm supp}\,\psi=L_0$, where $L_0$ is the 2-regular compact open subgroup of $F$ given by \eqref{eq-def-Lq} and Lemma \ref{lem-totdisconn-threesets}. Once the claim is established, we get the desired conclusion directly from Corollary \ref{cor-totdisconn-supp} and the condition $\|\psi\|_{L^2(F)}=1$.

For the claim we recall the fact
    $$(*)\;\; y \in L_q \Leftrightarrow q \pm 2^{-1}y \in {\rm supp}\,\psi.$$
We begin with $x\in L_0$, then we have $2x\in L_0 \Leftrightarrow \pm x \in {\rm supp}\,\psi$ by $(*)$ with $q=0$. This gives us the inclusion $L_0\subset {\rm supp}\,\psi$.
For the converse we consider $x\in {\rm supp}\,\psi$. Corollary \ref{cor-totdisconn-supp} says that ${\rm supp}\,\psi$ is balanced, then we have $2^{-1}x\in {\rm supp}\,\psi$ from the assumption $0\in {\rm supp}\,\psi$. Now we apply $(*)$ with $q=0$ and the fact that $L_0$ is a group to get  $2^{-1}x\pm 2^{-1}x\in {\rm supp}\,\psi$, which is equivalent to $x\in L_{2^{-1}x}$ by $(*)$ with $q=2^{-1}x$. Since $L_{2^{-1}x}$ is also a group by Lemma \ref{lem-totdisconn-threesets}, we have $2x\in L_{2^{-1}x}$ and therefore $-2^{-1}x=2^{-1}x-x\in {\rm supp}\,\psi$ by $(*)$ with $q=2^{-1}x$, which means that $x\in L_0$ by $(*)$ with $q=0$.
\end{proof}

We finally complete the proof of Theorem \ref{thm-Hudson-totdisconn}.

\begin{proof}[Proof of Theorem \ref{thm-Hudson-totdisconn}] $(2)\Rightarrow(1)$:
Starting from \eqref{eq-psi} of Lemma \ref{lem-totdisconn-Hudson2}, we have
    $$|\varphi_q(x)|=\mu_F(K)^{-1}1_{x_0+K}(q+2^{-1}x)1_{x_0+K}(q-2^{-1}x)=\mu_F(K)^{-1}1_{x_0+K}(q)1_{K}(x),$$
where we used the 2-regularity of $K$ in the last equality. Moreover, since $\varphi_q$ is continuous and positive definite on $K$, Proposition \ref{prop-posdef}(2) implies that
\begin{equation} \label{eq-totdisconn-varphi1}
    \varphi_q(x)=\mu_F(K)^{-1}1_{x_0+K}(q)1_K(x)\gamma_q(x)
\end{equation}
for some $\gamma_q\in \widehat{K}$. Therefore, we get the Wigner function
$$\mathcal{W}_{\psi}(q,p)=\widehat{\varphi_q}^F(p)=1_{x_0+K}(q)1_{K^{\perp}}(p-\widetilde{\gamma}_q),$$
where $\tilde{\gamma}_q\in \widehat{F}$ is any extension of $\gamma_q$. Here, we use the fact that characters on a closed subgroup can be extended to a character on the whole group \cite[Theorem 4.2.14]{ReiSte}. Now, by considering $\psi_0:=W(x_0,\widetilde{\gamma}_0)^*\psi$ combined with \eqref{eq-wignerclifford}, we may assume that $x_0=0$ and $\gamma_0\equiv 1$.

Going back to \eqref{eq-psi} we can write
    $$\psi(x)=\mu_F(K)^{-1/2}1_K(x)\alpha(x)$$
for some continuous function $\alpha$ on $K$ with $|\alpha|\equiv 1$, which gives us
\begin{equation}\label{eq-totdisconn-varphi2}
    \varphi_q(x)=\mu_F(K)^{-1}1_K(q)1_K(x)\alpha(q+2^{-1}x)\overline{\alpha(q-2^{-1}x)}, \; x\in F.
\end{equation}
Comparing \eqref{eq-totdisconn-varphi1} and \eqref{eq-totdisconn-varphi2} (under the condition $x_0=0$), we have
    \begin{equation}\label{eq-alpha-beta}
        \alpha(q+2^{-1}x)\overline{\alpha(q-2^{-1}x)}=\gamma_q(x),\;\;q,x\in K.
    \end{equation}
However, the condition $\gamma_0\equiv 1$ implies that $\alpha(2^{-1}x)=\alpha(-2^{-1}x)$ for all $x \in K$, which means $\alpha$ is symmetric thanks to 2-regularity of $K$. Therefore,
\begin{align*}
    \gamma_q(x)&=\alpha(2^{-1}x+q)\overline{\alpha(2^{-1}x-q)}\\
    &=\gamma_{2^{-1}x}(2q)=(\gamma_{2^{-1}x}(q))^2\\
    &=\left(\alpha(2^{-1}(x+q))\overline{\alpha(2^{-1}(x-q))}\right)^2\\
    &=\left(\alpha(2^{-1}(q+x))\overline{\alpha(2^{-1}(q-x))}\right)^2\\
    &=\gamma_x(q),\;\; q,x \in K.
\end{align*}
Consequently, we get a symmetric bicharacter $\beta: K\times K \to \tor,\; (q,x)\mapsto \gamma_q(x)$ introduced in Section \ref{sec-pureGaussian}. From the condition \eqref{eq-alpha-beta} we can easily see that $\alpha(x) = \beta(x,2^{-1}x)$, $x\in K$, which is the conclusion we wanted as in \eqref{eq-pure-Gaussian}.
\end{proof}

\begin{question}
Can we further generalize the Hudson theorem over 2-regular LCA group with compact open subgroups?
%\tcb{To be added, what makes hard to generalize the theorem.}
\end{question}

\begin{rem}
Note that the original Hudson's theorem~\cite{hudson1974wigner} and its higher dimensional generalization~\cite{soto1983wigner} do not assume the continuity of the vector state $\psi \in L^2(\Real^n)$. It can be deduced form the single assumption $\W_\psi \ge 0$ a.e..

On the other hand, a corresponding result on the angle-number system in 1-mode has been proved in \cite{rigas2010non}. A careful inspection of the proof reveals that an implicit assumption of the continuity of $\psi \in L^2(F)$ is made in \cite{rigas2010non}. It is not clear whether we could remove the continuity of $\psi \in L^2(F)$ from the assumption in both of the cases at the time of this writing.
\end{rem}

\section{Conclusion and Outlook}

In this paper we have established a framework to pursue Gaussian quantum information over general quantum kinematical systems (with finitely many degrees of freedom), and characterized Gaussian states over many systems of interest; 2-regular Weyl systems in particular. In addition to paving the way towards a general theory of Gaussian quantum channels (to appear \cite{GQIT2}), several natural questions concerning the structure and applications of Gaussian states remain open. For instance, separability versus positivity under partial transpose, purifications and optimization scenarios.

%({\bf Red part TO BE ERASED})
%{\color{red}

%\begin{itemize}
%    \item Entanglement of B-Gaussian states. In particular, \textbf{PPT=separable}?
    
%    \item (B-Gaussian) purification of B-Gaussian states.
    
%    \item Information quantities and distance measures (e.g. R\'enyi entropy, purity, fidelity, trace distance, Hilbert-Schumidt inner product...).
    
%    \item Optimality of entropy under certain constraints (e.g. see \cite[Lemma 12.25]{Hol} for \textbf{bosonic Gaussian states}, \cite[Proposition 2]{Z2} for \textbf{p-adic Gaussian states}) $\to$ \textbf{We can add this now...?}
    
%    \item Gaussian quantum protocols after in paper II (e.g. teleportation...).
%\end{itemize}

%}

\subsection*{Acknowledgements} HHL and SJP's research was supported by the Basic Science Research Program through the National Research Foundation of Korea (NRF) Grant NRF-2017R1E1A1A03070510 and the National Research Foundation of Korea (NRF) Grant funded by the Korean Government (MSIT) (Grant No.2017R1A5A1015626). JC's research was partially supported by the NSERC Discovery Grant RGPIN-2017-06275. JC would like to acknowledge helpful discussions with Zachary Zanussi. S.-G. Youn was funded by Samsung Science and Technology Foundation under Project Number SSTF-BA2002-01 and
by the National Research Foundation of Korea (NRF) grant funded by the Korea government (MSIT) (No. 2020R1C1C1A01009681).

\subsection*{Data Availability}
Data sharing not applicable to this article as no datasets were generated or analysed during the current study.

\end{document}